\documentclass[pra,preprint,showpacs]{revtex4}
\usepackage[centertags]{amsmath}
\usepackage{amsfonts}
\usepackage{enumerate}
\usepackage{amssymb}
\usepackage{amsthm} 
\usepackage{newlfont}
\usepackage{epsfig}
\usepackage{amscd}
\usepackage{graphicx}
\usepackage{epsfig}
\usepackage{footnote}
\usepackage{lipsum}
\usepackage{color}
\usepackage{xcolor}
\usepackage{graphicx}
\usepackage{multirow}
\usepackage{subfig}
\usepackage{amscd}

\newcommand{\beq}{\begin{equation}}
\newcommand{\eeq}{\end{equation}}
\newcommand{\ba}{\begin{array}}
\newcommand{\ea}{\end{array}}
\newcommand{\bea}{\begin{eqnarray}}
\newcommand{\eea}{\end{eqnarray}}
\begin{document}

\begin{center}
{\large \sc \bf {Entanglement in block-scalable and   block-scaled states}
}

\vskip 15pt

{\large 
G.A.Bochkin, S.I.Doronin and A.I.Zenchuk 
}

\vskip 8pt

{\it Institute of Problems of Chemical Physics, RAS,
Chernogolovka, Moscow reg., 142432, Russia}

\end{center}
\today

\begin{abstract}{
There is  a special class of so-called block-scalable initial states of the sender whose transfer to the receiver through the spin chain results in multiplying {their} MQ-coherence matrices by  scalar factors (block-scaled receiver's states). We  study the entanglement in block-scalable and block-scaled states and show that, generically, the entanglement in block scaled states is less than the entanglement in the corresponding block-scalable states, although  the balance  between these entanglements depends on particular values of parameters characterizing these states.  Using the perturbations of the sender's block-scalable  initial states we show that generically the entanglement in the block-scalable states is bigger, while the entanglement in the block-scaled states is less than the entanglements in the appropriate states from their neighborhoods.  A short  chain of 6  and 
a long chain of 42 spin-1/2 particles are considered as examples.}
\end{abstract}

\maketitle

\section{Introduction}
\label{Section:Introduction}

The {quantum}  state transfer \cite{Bose}, either perfect \cite{CDEL,ACDE,KS} or high-probability \cite{GKMT,WLKGGB}, is   {an attractive  field of} quantum information theory.  The problem of remote state creation  symbolizes  further progress in {development of this research area} and was first {implemented} in
photon systems \cite{PBGWK2,PBGWK,XLYG}. Unlike the state transfer, when creating the required  state we obtain a density matrix at the receiver site  which completely differs from the   state initially prepared at the sender. 
{ Nevertheless,  the link between the sender's initial state and the appropriate receiver's state 
can be completely understood \cite{BZ_2015} in the case of one-qubit state creation.  On the contrary,}  the situation becomes much more complicated in the case of multi-qubit state creation  because of {the} increasing number of creatable parameters. Therefore  the control of multi-qubit creatable states becomes more { intricate}. 

{ In this respect we shall note that remote preparation of particular multi-qubit states was studied in photon systems  
\cite{WSZXXJ,WSLZKYWJZ,WSMXZKL}. The
protocol proposed in those papers is based on two-qubit entangled states and combines a unitary transformation with measurements \cite{SLRV,VLUKED} on the sender side.  A classical communication channel is required in that case as well. We also note that the sender possesses
the complete information about the state to be created at the receiver side, unlike the teleportation protocol  \cite{BBCJPW,BPMEWZ,BBMHP}.}

In our paper, we are interested in a protocol for creating  {such}  families of states  that are well-described deformations of the initial sender's states. 
In this case, the  straightforward  control over the creatable states can be {simply} established. 
In Ref.\cite{BFZ_Arch2018}, so called block-scaled states {of the receiver} have been introduced as such deformations. { These} states  differ from the initial sender's state by  scale factors in front of  certain blocks, which are the multiple-quantum (MQ) coherence matrices. {The advantage of such a block-representation is that each MQ-coherence matrix evolve independently} if  certain conditions are satisfied {\cite{FZ_2017}}. We  recall that 
the $k$-order  coherence matrix consists of those elements of a  density matrix that are responsible for the state-transitions changing the $z$-projection
$I_z$ of the total spin momentum by $k$. The sender's initial state  which can be transferred to the receiver as a block-scaled one is called the block-scalable state.  
We emphasize that the  MQ coherence matrices in the sender's {block-scalable} initial state  can not be arbitrary. {They} must have a prescribed structure {and this structure admits only one scalar parameter encoded} into the pair of $\pm n$-order  coherence matrices ($n>0$). Thus,  {that} protocol for creating the {$M$-qubit} block-scaled states allows us to construct a family of states rather than  particular {states}, and this family is parametrized by { a set of $M$} free parameters { $c^{(1)}$ $\dots$ $c^{(M)}$} transferred by the { $\pm 1$- $\dots$ $\pm M$-order} coherence matrices. {
We emphasize that the sender doesn't have to know the values of the
parameters $c^{(i)}$ to be transferred to the receiver side. There is also no
needs for a classical communication channel. Therefore, it is better  to
consider the protocol of block-scaled state creation as a further development
of the state transfer protocol rather than a development of the
teleportation protocol.}

In this paper, we study  { the entanglements  in {2-qubit} block-scalable and corresponding block-scaled states as  functions of  the  parameters $c^{(1)}$ and $c^{(2)}$ {transferred by the $\pm 1$- and $\pm 2$-order coherence matrices}. Generically, the entanglement in the block-scaled states $C(\rho^{(R)})$  is less than the entanglement in the block-scalable states $C(\rho^{(S)})$. However, depending on the orders of the {MQ}-coherence matrices involved in the  process, the entanglement $C(\rho^{(S)})$  can be either greater or less than the appropriate  entanglement $C(\rho^{(R)})$. Thus, if only the zero- and $\pm 2$-order coherences are involved then  $C(\rho^{(R)})< C(\rho^{(S)})$. If the zero- and $\pm 1$-order coherences are involved then the situation is generally opposite,  $C(\rho^{(R)}) > C(\rho^{(S)})$ and $C(\rho^{(R)})$ is always bigger than zero.  If all the above  coherence matrices are involved, then  the balance between $C(\rho^{(R)})$ and  $C(\rho^{(S)})$  depends on the values of the parameters $c^{(1)}$ and $c^{(2)}$, although the maximal value of $C(\rho^{(R)})$ is significantly smaller than that of $C(\rho^{(S)})$.}

{Then, we compare the entanglement in block-scalable (block-scaled) states with the entanglement in states from the close and remote neighborhoods of these states.
For this purpose  we  study}  the perturbations of the block-scalable states. The {influence of  perturbations on the entanglement}  depends on  {its amplitude} and on the particular values of the transferred parameters $c^{(1)}$ and $c^{(2)}$. But generically, the maximal value of entanglement { first decreases with an increase in $\varepsilon$ till $\varepsilon \sim 0.1$, and then it decreases with the further  increase in $\varepsilon$. The general behavior of $C(\rho^{(R)})$ is opposite. It, first, decreases with an increase in $\varepsilon$ till $\varepsilon\sim 0.02$, and then it increases with the further increase in $\varepsilon$. Some exceptions from this rule are discussed in Sec.\ref{Section:num2}}.  { In other words, in general, the map (block-scalable states) $\to$ (block-scaled states) reduces the quantum correlations more significantly than the map of states from the neighborhood.}

The paper is organized as follows. The block-scaled state formation is reviewed in Sec.\ref{Section:block}. In Sec.\ref{Section:conc}, the entanglement in the block-scalable and block-scaled states is {studied} as a function of the transferred free parameters {$c^{(1)}$ and $c^{(2)}$}. Effect of perturbation of the block-scalable states on the entanglement is {considered} in Sec.\ref{Section:pert}. The paper is concluded with Sec.\ref{Section:conclusions}. {The density matrices}  of the {2-qubit} block-scalable { and block-scaled} states {found in \cite{BFZ_Arch2018}} for the chains of { $N=6$ and $42$ spins} are given in  Appendix, Sec.\ref{Section:AppendixA}.

\section{Block-scaled state formation}
\label{Section:block}
{To study the entanglement in block-scalable and corresponding block-scaled states we turn to the  model considered}  in Ref. \cite{BFZ_Arch2018}. That model is a  tripartite  communication line consisting of {the}  two-qubit sender $S$, {the} transmission line $TL$ and {the} two-qubit receiver $R$. { The initial state of  the whole system is the tensor product one,} 
\begin{eqnarray}\label{in2}
\rho(0)=\rho^{(S)}(0) \otimes \rho^{(TL,R)}(0),
\end{eqnarray}
where $\rho^{(S)}(0)$ is an arbitrary  initial state of the sender $S$, and 
\begin{eqnarray}
 \label{inTLB2}
\rho^{(TL,R)} &=&\frac{e^{bI^{(TL;R)}_z}}{\left(2 \cosh\frac{b}{2}\right)^{N-2}}
 \end{eqnarray}
 is the thermal equilibrium initial state of the subsystem $TL\cup R$ with $b=\frac{\hbar \omega_0}{kT}$, $\hbar$ is the Planck constant,
 $k$ is the Boltzmann constant,  $\omega_0$ is the Larmor frequency, $T$ is the temperature and $I^{(TL;R)}_z$ is the $z$-projection of the total spin-momentum of the subsystem  $TL\cup R$.
 Here, both the sender and receiver are two-qubit subsystems. 
The  evolution is described by the Liouville equation
\begin{eqnarray}
\rho(t)= V(t) \rho(0) V^+(t),\;\;\;V(t)=e^{-i H t},
\end{eqnarray}
where $H$ is  the $XX$-Hamiltonian 
\begin{eqnarray}\label{XY}
H=\sum_{i=1}^{N-1} D (I_{ix}I_{(i+1)x} +I_{iy}I_{(i+1)y}),
\end{eqnarray}
$D$ is {the} coupling constant and $I_{i\alpha}$, $\alpha=x,y,z$, is {the projection operator} of the $i$th spin momentum into the $\alpha$-axis.
The density matrix of the receiver at the time instant $t$ can be   reduced from the whole density matrix $\rho(t)$ as:
\begin{eqnarray}
 \rho^{(R)}(t)={\mbox Tr}_{S,TL}\rho(t).
 \end{eqnarray}
Both  density matrices $\rho^{(S)}(0)$ and $\rho^{(R)}(t) $  can be represented as the sums of MQ coherence matrices:
\begin{eqnarray}
\rho^{(S)}(0) = \sum_{i=-2}^2\rho^{(S;i)}(0),\;\;\;\rho^{(R)}(t) = \sum_{i=-2}^2\rho^{(R;i)}(t).
\end{eqnarray}
{An} important feature of the Hamiltonian (\ref{XY}) is that it commutes with $I_z$: $[H,I_z]=0$.
This fact ensures the independent evolution of the  multiple-quantum coherences \cite{FZ_2017} and
 together with the chosen initial state  (\ref{in2}), (\ref{inTLB2}) {allows to describe the state transfer as the map} 
 \begin{eqnarray}\label{map}
 \rho^{(S;\pm n)}(0) \to \rho^{(R;\pm n)}(t),\;\;\;n=0,\pm1,\pm2. 
 \end{eqnarray}
Obviously,  all elements of $\rho^{(S;\pm n)}(0)$ are mixed in $\rho^{(R;\pm n)}(t)$ in general unless 
the matrices $\rho^{(S;\pm n)}(0)$ have the special structures found in Ref.\cite{BFZ_Arch2018}:
\begin{eqnarray}
&&
\rho^{(S;\pm n)}(0) = c^{(n)} \rho^{(X;\pm n)}(0),{\;\;n>0,}\\\nonumber
&&
\rho^{(S;0)} = e^{(4)} + \tilde \rho^{(X;0)}
,\;\;\;e^{(4)}  = {\mbox{diag}} (0,0,0,1).
\end{eqnarray}
Here, $c^{(n)}$ are constant parameters which are  real { positive} for simplicity  (therefore $c^{(-n)}=c^{(n}$), while $\rho^{(X;\pm n)}(0)$ and $\tilde \rho^{(X;0)}(0)$ are the constant matrices uniquely fixed by the optimization procedure \cite{BFZ_Arch2018}\footnote{Explicit forms for the matrices 
$\rho^{(X;\pm 1)}(0)$ and $\tilde \rho^{(X;0)}(0)$ for the  communications lines of $N=6$ and $42$ nodes and different structures of the transferable matrix   at the appropriate time instant 
are given in the Appendix, Sec.\ref{Section:AppendixA})}.
Then,  map (\ref{map}) at the properly chosen time instant $t$ can be written as:
\begin{eqnarray}\label{map2}
&&
c^{(n)} \rho^{(X;\pm n)}(0) \to \lambda^{(\pm n)}{(t)} c^{(n)} \rho^{(X;\pm n)}(0),\;\;n= 1, 2,\\\nonumber
&&
 \tilde \rho^{(X;0)}(0) \to \lambda^{(0)}{(t)} \tilde  \rho^{(X;0)}(0),
\end{eqnarray}
where $\lambda^{(i)}$, $i=0,\pm 1,\pm 2$, are some scale  factors which { are considered to be real positive for simplicity} and therefore $\lambda^{(n)} =\lambda^{(-n)}$. 
It is remarkable that $\lambda^{(0)}$ can be equal to one, i.e., {the} zero-order coherence can be mapped without any deformation
(perfect transfer),
while other scales {turned out} to be less then one (compressive map):
$\lambda^{(n)}<1$, $n> 0$.  
{The initial sender's state and the appropriate receiver's state related to each other by  map (\ref{map2}) are referred to as, respectively, the block-scalable and block-scaled states  \cite{BFZ_Arch2018}}. 
{ Thus, the parameters $c^{(i)}$, $i=1,2$, characterize the space of the sender's block-scalable states and the parameters $\lambda^{(i)}$, $i=1,2$, show  the compression rate of the corresponding {creatable} space of the receiver's block-scaled states. The parameter $\lambda^{(0)}$ describing the compression  of the zero-order coherence matrix can serve to maximize the creatable $(c^{(1)},c^{(2)})$-space \cite{BFZ_Arch2018}. All the parameters $c^{(i)}$ and $\lambda^{(i)}$ depend on the chain length and the system  Hamiltonian.}

In ref.\cite{BFZ_Arch2018}, { the block-scaled state transfer was studied in the two settings for $\lambda^{(0)}$:} $\lambda^{(0)}=1$   and $\lambda^{(0)}=\lambda^{(0)}_{opt} $ 
($\lambda^{(0)}_{opt}$  is { the value of $\lambda^{(0)}$ that maximizes} the creatable region of the receiver state space on the plane of the  parameters $(c^{(1)}, c^{(2)})$). However, the direct analysis shows that  the former case yields zero entanglement in both $\rho^{(S)}$ and $\rho^{(R)}$. This prompts us to { assume} that the requirement of block-scaled state transfer reduces quantum correlations in {receiver's} states, which is generally justified below. {Thus}, only the case $\lambda^{(0)}=\lambda^{(0)}_{opt} $ is considered hereafter.

{According to Ref.\cite{BFZ_Arch2018}, we consider the following sets of  constraints on the { transferred} parameters $c^{(i)}$ and { on the scale factors} $\lambda^{(i)}$, $i=1,2$.}

\noindent
{{\bf Case I:}} $\lambda^{1}\neq \lambda^{2}$,
$c^{(1)}=0$, {$c^{(2)} \ge 0$}, the $\pm$1-order coherence matrices are absent, $0\le c^{(2)}\le c^{(2)}_{max}$.

\noindent
{{\bf Case II:}}
$\lambda^{1}\neq \lambda^{2}$,
{ $c^{(1)}\ge 0$}, $c^{(2)} = 0$, the $\pm$2-order coherence matrices are absent, $0\le c^{(1)}\le c^{(1)}_{max}$.

\noindent
{{\bf Case III:}} $\lambda^{1}\neq \lambda^{2}$,
{ $c^{(1)}\ge 0$, $c^{(2)} \ge 0$},  all {the}  coherence matrices are present, the allowed domain on the plane $(c^{(1)},c^{(2)})$ is {a quarter} of an  ellipse-like region with the semi-axes $c^{(1)}_{max}$ and $c^{(2)}_{max}$.

\noindent
{{\bf Case IV:}}
$\lambda^{1} = \lambda^{2}$, { $c^{(1)}\ge 0$, $c^{(2)} \ge 0$}, uniform scaling of the higher order coherence matrices.  Similar to the previous case, the allowed domain on the plane $(c^{(1)},c^{(2)})$ is {a quarter} of an  ellipse-like region with the semi-axes $c^{(1)}_{max}$ and $c^{(2)}_{max}$.

Here the parameters  $c^{(i)}_{max}$, {$i=1,2$,} depend on the length of the communication line.

\section{Entanglement in block-scalable and block-scaled states}
\label{Section:conc}

We study the entanglement using the Wootters criterion \cite{HW,Wootters} in terms of concurrence 
according to the formula
\begin{eqnarray}\label{C}
C=\max(0,2\lambda_{max} -\sum_{i=1}^4\lambda_i), \;\;\lambda_{max}=\max(\lambda_1,\dots,\lambda_4),
\end{eqnarray}
where $\lambda_i$ are the eigenvalues of the matrix
\begin{eqnarray}\label{trho}
\tilde\rho =\sqrt{\rho (\sigma_y\otimes \sigma_y)(\rho)^* (\sigma_y\otimes \sigma_y) }.
\end{eqnarray}
Here {$\rho$ is the density matrix of either the sender's or receiver's state}, $*$ means {the} complex conjugate and $\sigma_y$ is the Pauli matrix.

According to Sec.\ref{Section:block},
{ the density matrices in our model have the following structure:}
\begin{eqnarray}\label{rhoSB}
 \rho^{(S)} &=& e^{(4)} + \tilde \rho^{(X;0)}+c^{(1)} \Big(\rho^{(X;1)}+ \rho^{(X;-1)}\Big) +  
c^{(2)} \Big(\rho^{(X;2)} +  \rho^{(X;-2)}\Big),\\\label{rhoRB}
\rho^{(R)} &=& e^{(4)} + \lambda^{(0)} \tilde \rho^{(X;0)}+ \lambda^{(1)}c^{(1)} \Big( \rho^{(X;1)}+
 \rho^{(X;-1)}\Big) +
\lambda^{(2)} c^{(2)} \Big(  \rho^{(X;2)} + \rho^{(X;-2)}\Big).
\end{eqnarray}

\subsection{Systems of $N=6$ and $N=42$ spins
}
\label{Section:num}
Now, we describe the entanglement in four cases listed in the end of Sec.\ref{Section:block}.
\subsubsection{{ \bf Case I:} $\lambda^{1}\neq \lambda^{2}$,
$c^{(1)}=0$, {$c^{(2)}\ge 0$}.} 
\label{Section:numCase1}
In this case the non-zero concurrence is a linear function of $c^{(2)}$, which 
can be  proved as follows.
The density matrices $\rho^{(R)}$ and $\rho^{(S)}$ have the following general  structure (see Appendix, eqs.(\ref{case11}), (\ref{case12})):
\begin{eqnarray}
\rho=\left(
\begin{array}{cccc}
a_{11}&0&0& s\cr
0& a_{22}&a_{23} i &0\cr
0& -a_{23} i &a_{33} &0\cr
s&0&0&a_{44}
\end{array}
\right),\;\;\sum_{i=1}^4 a_{ii}=1,
\end{eqnarray}
where $a_{ij}$ are {the} real constants,  $s=c^{(2)}$ for $\rho^{(S)}$ and $s=\lambda^{(2)} c^{(2)}$ for $\rho^{(R)}$.
The eigenvalues of $\tilde \rho$ (\ref{trho}) read
\begin{eqnarray}
\lambda_{1,2}=|a_{23}\pm  \sqrt{a_{22} a_{33}}|, \;\;\;\lambda_{3,4}=|s \pm \sqrt{a_{11}a_{44}}|,
\end{eqnarray}
which are linear functions of $s$. Therefore, the concurrence (\ref{C}) is either constant independent on $c^{(2)}$ (zero in particular)  or a linear function of $s$. $\Box$

\begin{figure*}
\subfloat[]{\includegraphics[scale=0.5,angle=0]{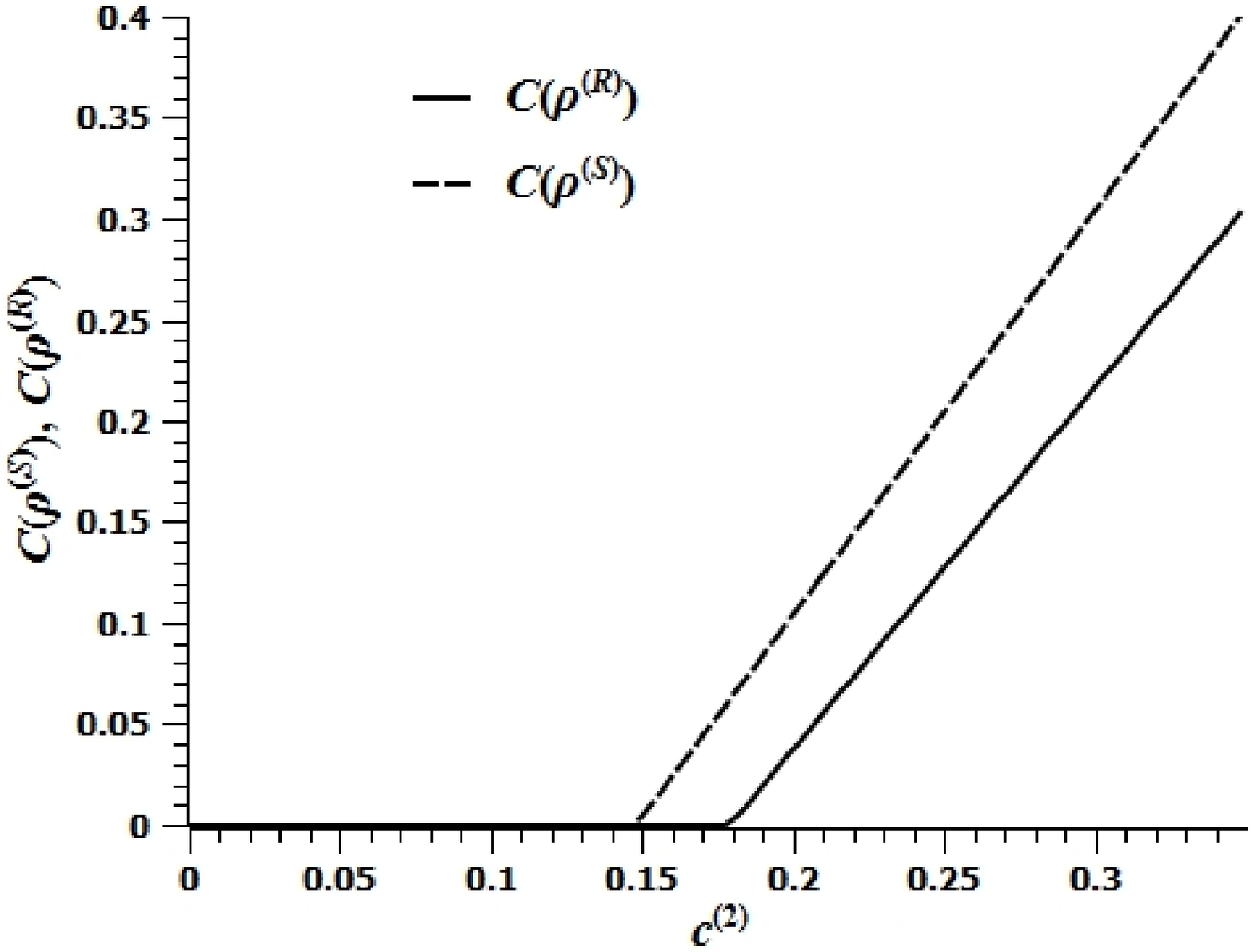
}}
\subfloat[]{\includegraphics[scale=0.5,angle=0]{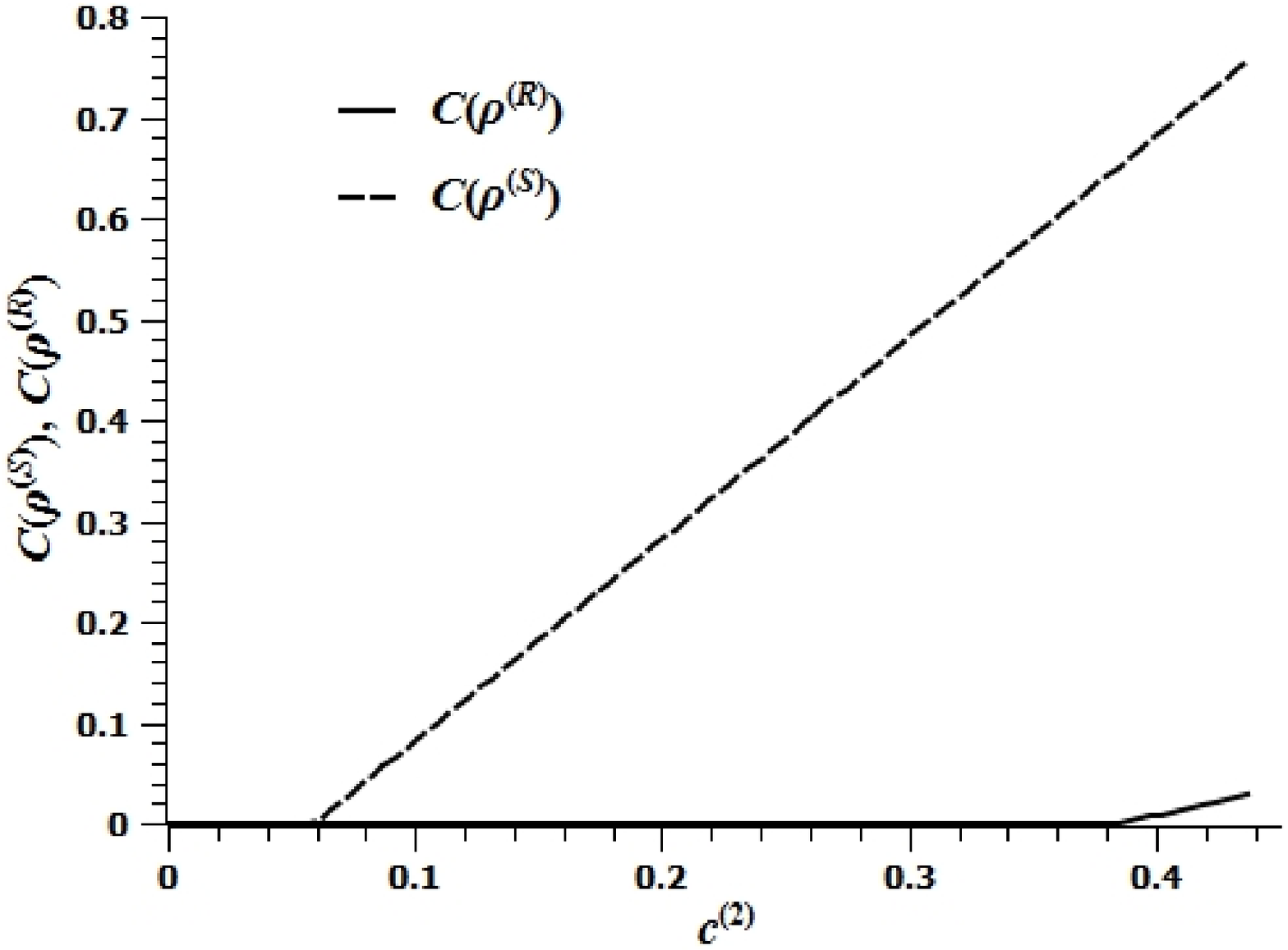
}}
\caption{{Case I:} 
$\lambda^{(1)}\neq\lambda^{(2)}$, $c^{(1)}=0$, {$c^{(2)}\ge 0$}. Concurrences  $C(\rho^{(S)})$ and $C(\rho^{(R)})$ as functions of $c^{(2)}$. 
(a) $N=6$, $c^{(2)}_{max}=0.3479 $; (b) $N=42$, $c^{(2)}_{max}=0.4372$. 
}
  \label{Fig:c1Eq0} 
\end{figure*}

Moreover, in both considered cases $N=6$ and $N=42$ {we have}
\begin{eqnarray}\label{a}
\sqrt{a_{11}a_{44}}>a_{23} + \sqrt{a_{22} a_{33}},
\end{eqnarray}
so that the concurrence reads
\begin{eqnarray}\label{Cl}
C=\max(0, 2 s - \lambda_1-\lambda_2),
\end{eqnarray}
which is zero for $s\le (\lambda_1+\lambda_2)/2$ and agrees with Fig.\ref{Fig:c1Eq0}. Of course, $\lambda_i$ in (\ref{Cl}) are different for the sender's and receiver's states which we label by the superscripts, respectively, $(S)$ and $(R)$. Thus, there is a critical value $c^{(2)}_{S;cr} = \frac{\lambda_1^{(S)}+\lambda_2^{(S)}}{2}$ and the appropriate critical value $c^{(2)}_{R;cr}= \frac{\lambda_1^{(R)}+\lambda_2^{(R)}}{2} > c^{(2)}_{S;cr} $ such that $C(\rho^{(S)})=0$ if $c^{(2)} <c^{(2)}_{S;cr}$ and 
$C(\rho^{(R)})=0$ if $c^{(2)} <c^{(2)}_{R;cr}$.

Naturally, the concurrence in $\rho^{(R)}$ is bigger for { the 6-spin chain than for the 42-spin one}, showing that the entanglement   decays in long  chains.  {On the contrary,} the maximal concurrence in $\rho^{(S)}$ (corresponding  to 
$c^{(2)}=c^{(2)}_{max}$)   for $N=42$ is about twice { as big as that}
for $N=6$. 
{Thus, the concurrence in the block-scalable states increases with $N$, unlike  the concurrence in the block-scaled states.}

\subsubsection{{ \bf Case II:} $\lambda^{1}\neq \lambda^{2}$,
{$c^{(1)}\ge 0$}, $c^{(2)} = 0$.} 
\begin{figure*}
\subfloat[]{\includegraphics[scale=0.5,angle=0]{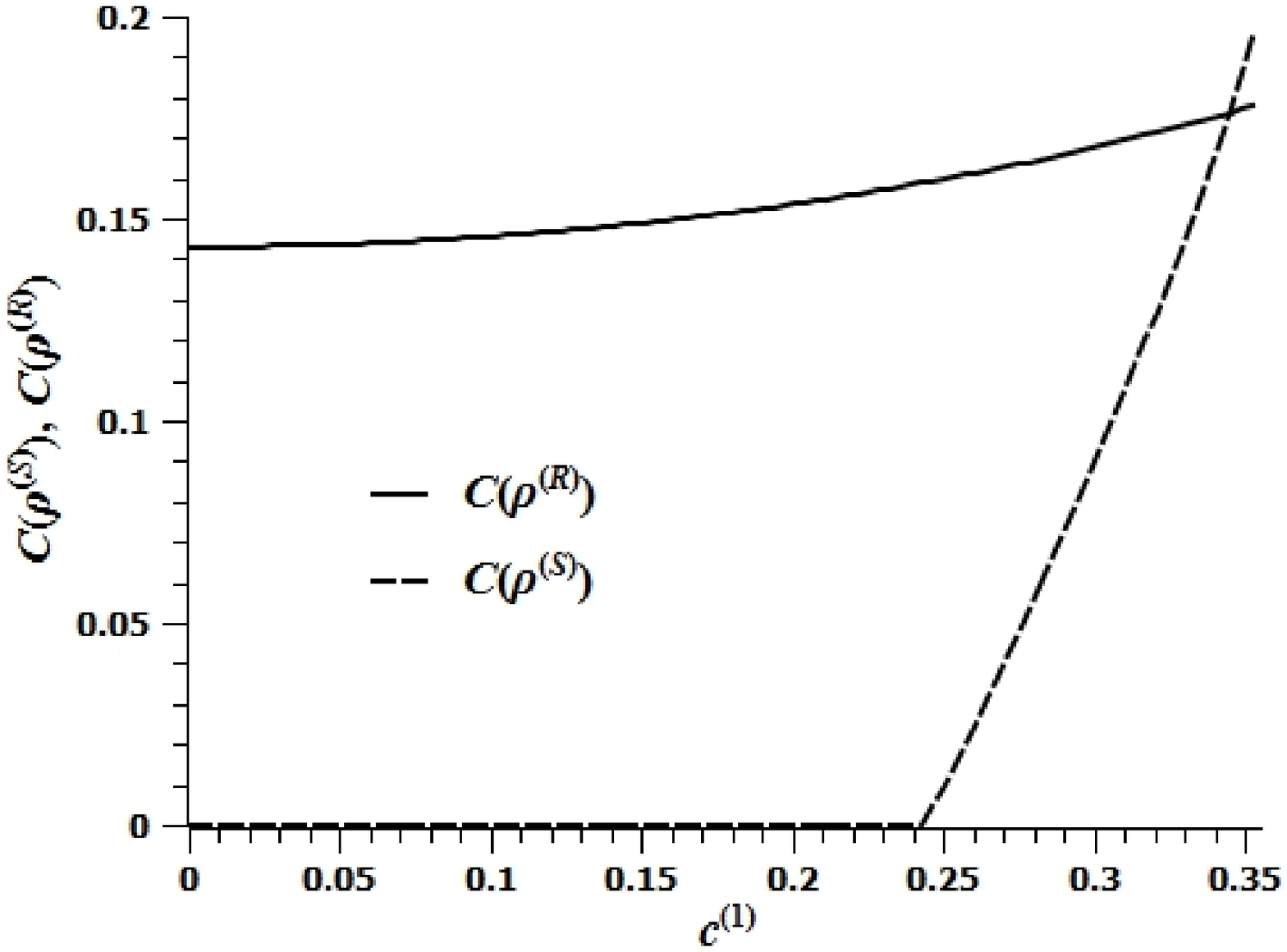
}}
\subfloat[]{\includegraphics[scale=0.5,angle=0]{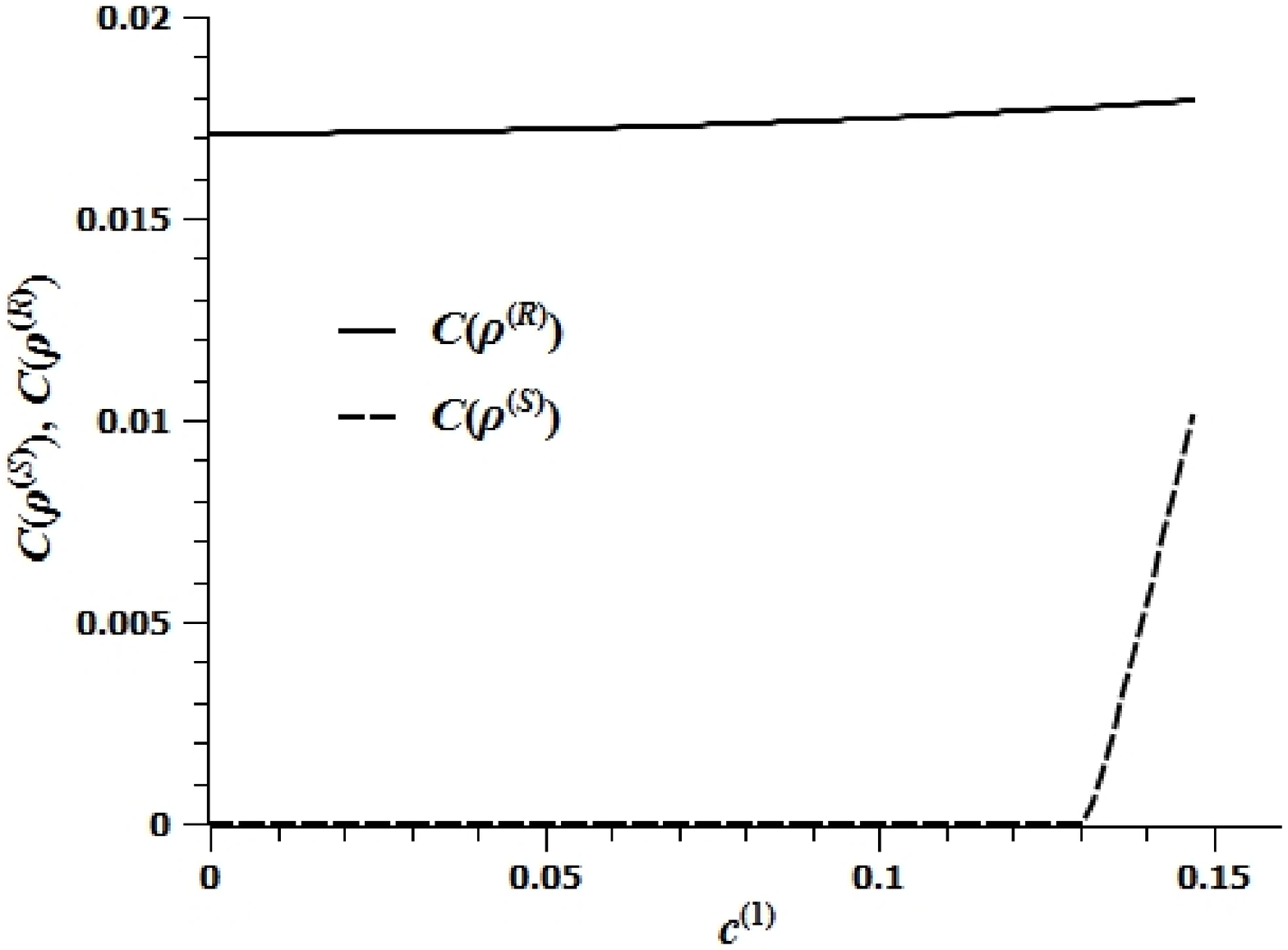
}}
\caption{{Case II:} $\lambda^{(1)}\neq\lambda^{(2)}$, {$c^{(1)}\ge 0$}, $c^{(2)}= 0$. Concurrences  $C(\rho^{(S)})$ and $C(\rho^{(R)})$ as functions of $c^{(1)}$.
(a) $N=6$, $c^{(1)}_{max}=0.3524$; (b) $N=42$, $c^{(1)}_{max}=0.1469$. 
}
  \label{Fig:c2Eq0} 
\end{figure*}

{ In this case, the
behavior of 
concurrence significantly differs from Case I. First of all, the concurrence is nonlinear function of
 $c^{(1)}$, which is confirmed by 
 Fig. \ref{Fig:c2Eq0}. {Next, the concurrence of the sender's state} $C(\rho^{(S)})=0$ if $c^{(1)} <c^{(1)}_{S;cr}$, while the
 concurrence of {the receiver's state} $\rho^{(R)}$ is non-zero for all $c^{(1)}$. {
 In addition, 
 the concurrence in $\rho^{(S)}$  {reaches larger values} in  the shorter chain ($N=6$, Fig.\ref{Fig:c2Eq0}a) than in the longer one ($N=42$, Fig.\ref{Fig:c2Eq0}b). Finally, $C(\rho^{(R)})>C(\rho^{(S)})$  over the allowed domain of $c^{(1)}$ except the case of the short chain  with $c^{(1)}$ approaching $c^{(1)}_{max}$ as shown in Fig.\ref{Fig:c2Eq0}a.}}

\subsubsection{{ \bf Case III:} $\lambda^{1}\neq \lambda^{2}$,
{$c^{(1)}\ge 0$, $c^{(2)} \ge 0$}.} 
\label{Section:CaseIII}
\begin{figure*}
\subfloat[]{\includegraphics[scale=0.6,angle=0]{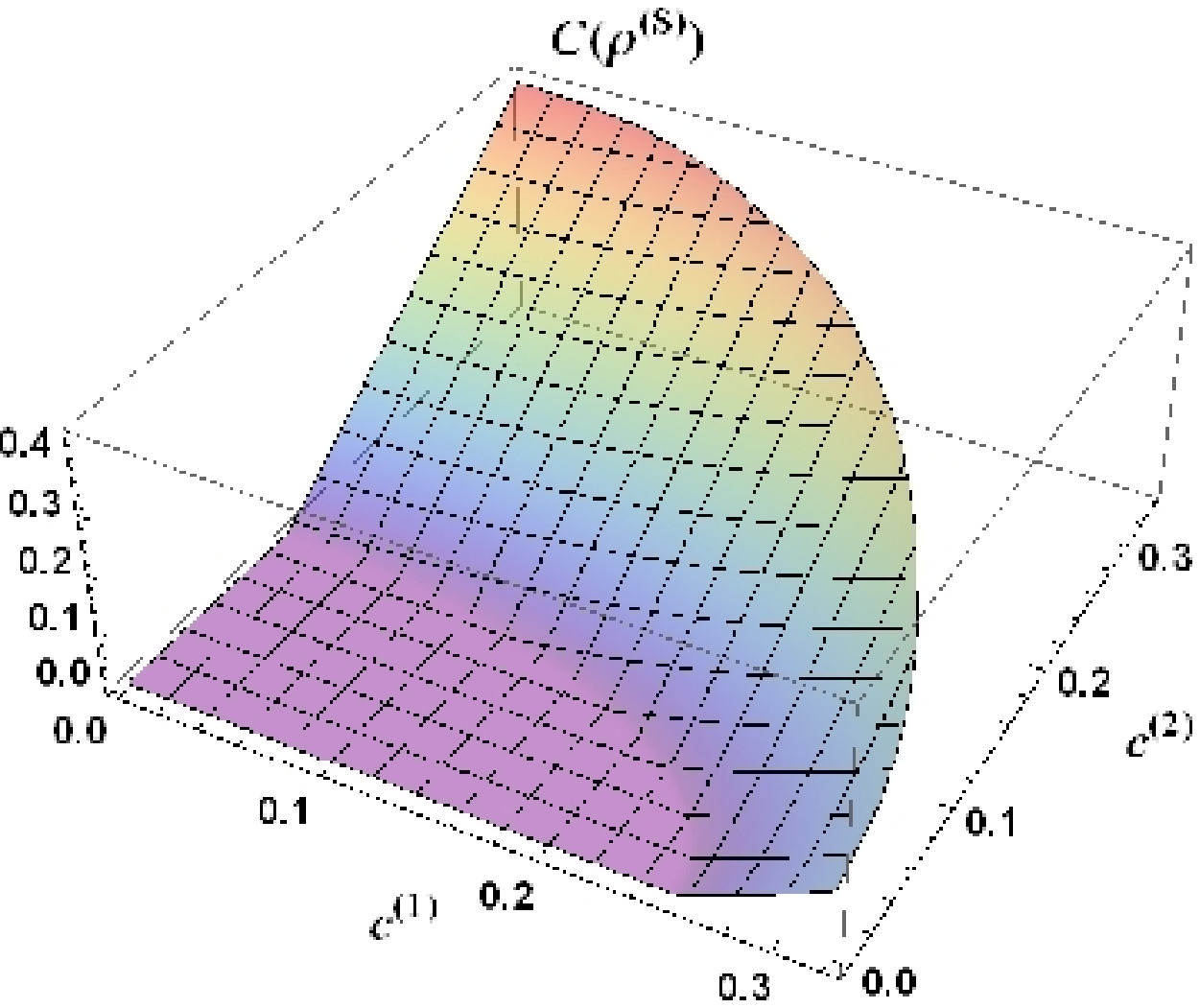
}}
\subfloat[]{\includegraphics[scale=0.6,angle=0]{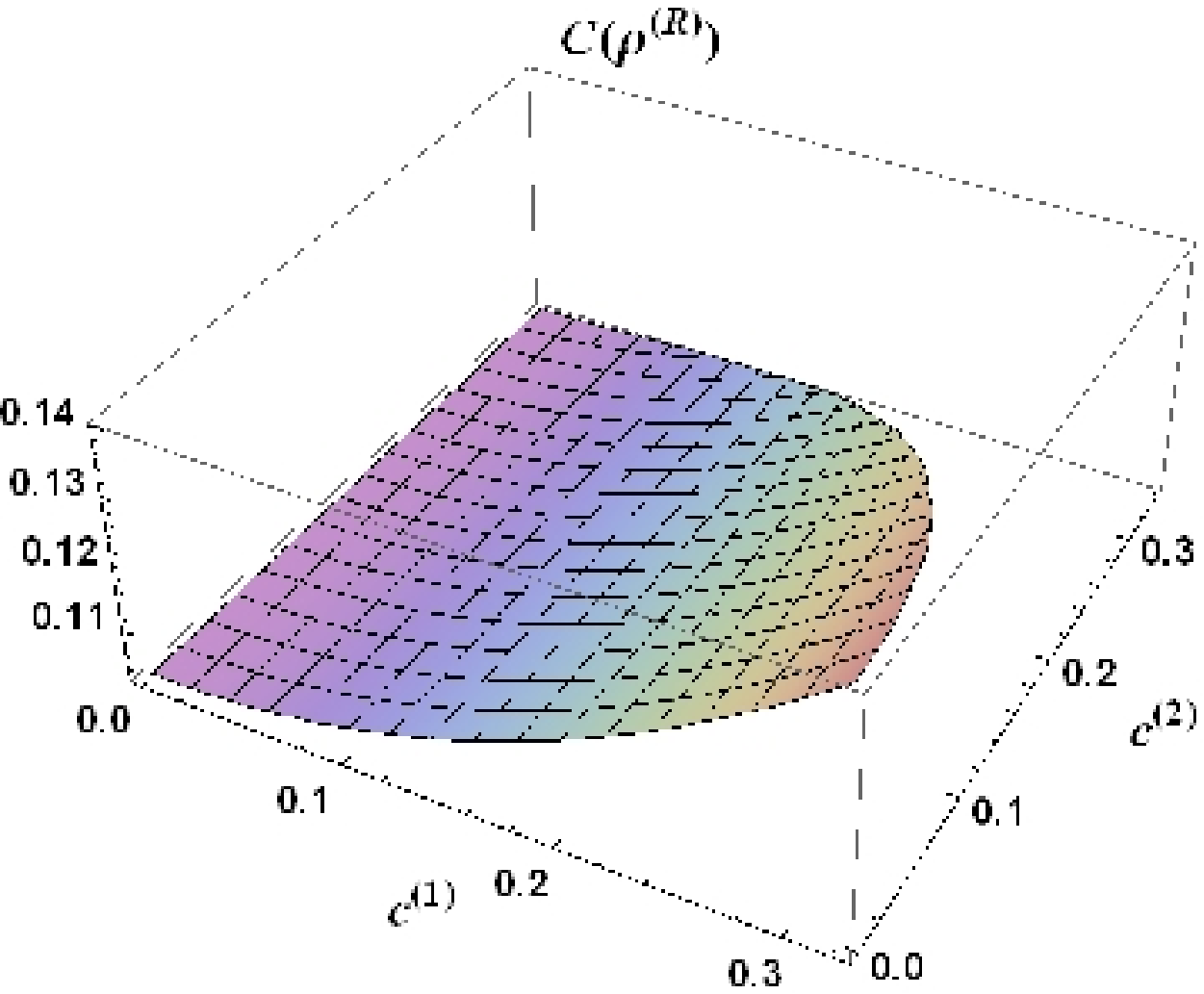
}}\\
\subfloat[]{\includegraphics[scale=0.6,angle=0]{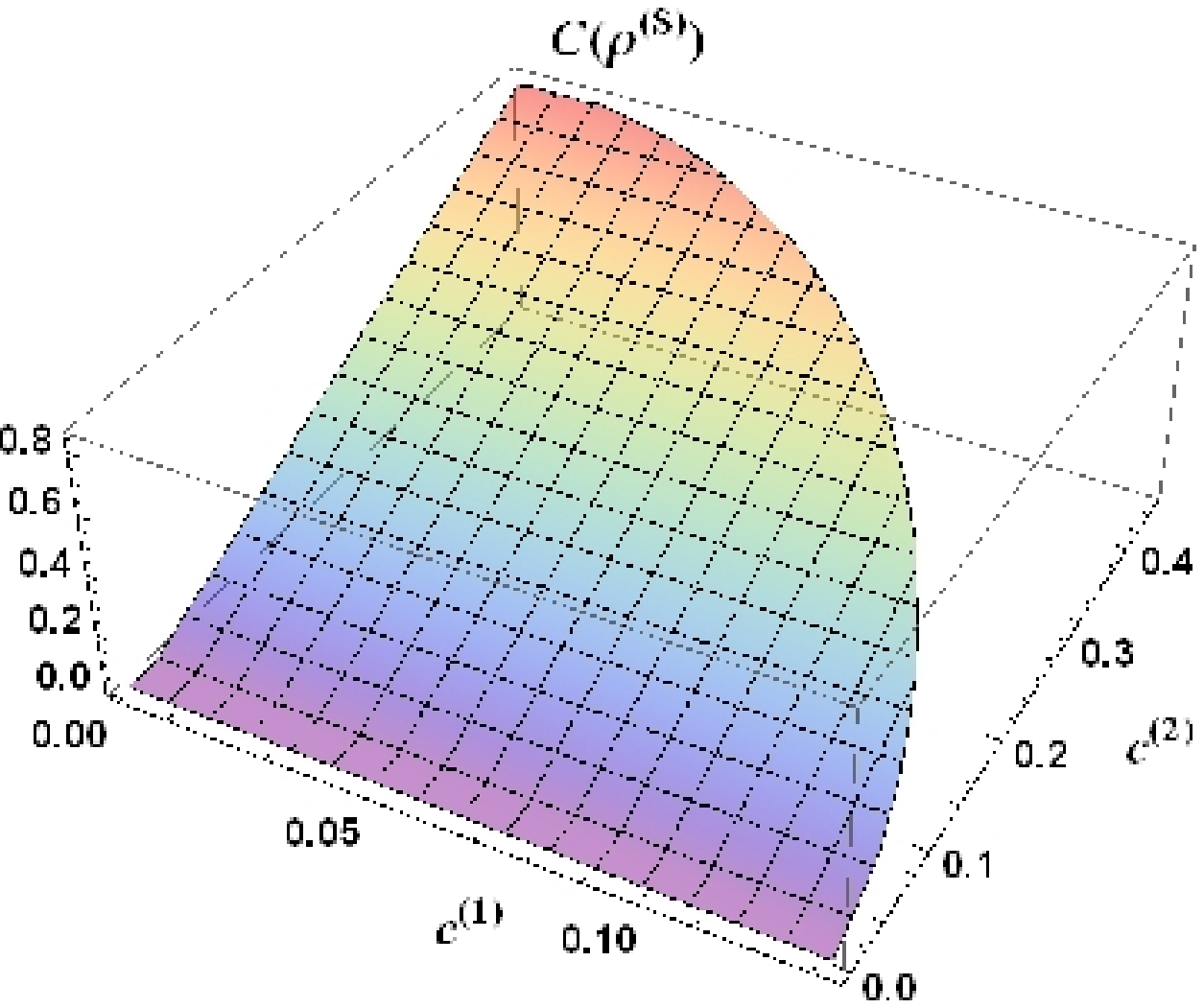
}}
\subfloat[]{\includegraphics[scale=0.6,angle=0]{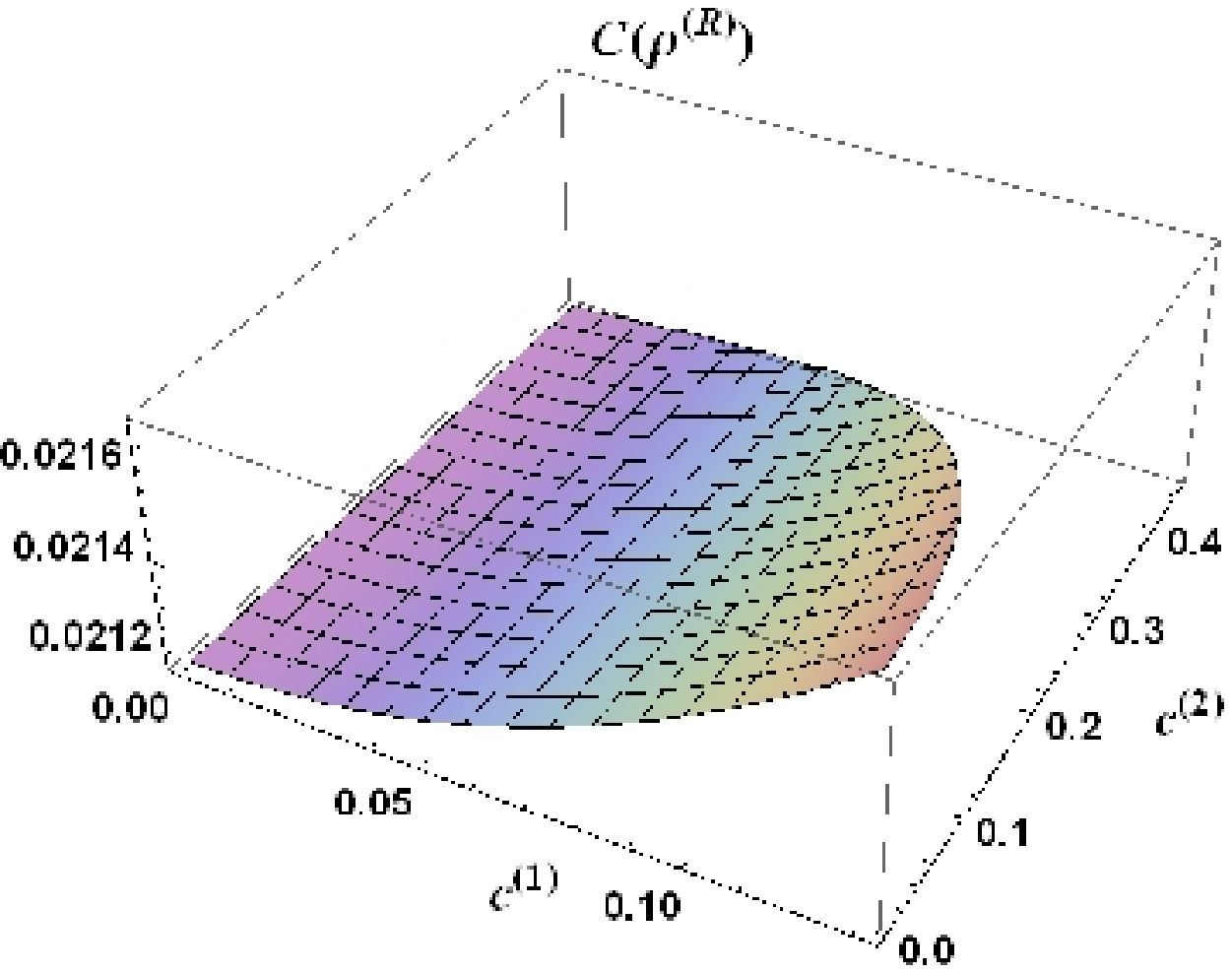
}}
\caption{{Case III:} $\lambda^{(1)}\neq\lambda^{(2)}$, {$c^{(1)}\ge 0$, $c^{(2)}\ge 0$}. Concurrences $C(\rho^{(S)})$ and $C(\rho^{(R)})$ as functions of $c^{(1)}$
and $c^{(2)}$  over the quarter of the ellipse-like domain on the plane $(c^{(1)},c^{(2)})$. { (a,b)  $N=6$,  the semi-axes 
$c^{(1)}_{max}=0.3216$ and $c^{(2)}_{max}=0.3369$; (a) $C(\rho^{(S)})$; (b)  $C(\rho^{(R)})$. (c,d) $N=42$, the semi-axes 
$c^{(1)}_{max}=0.1322$ and $c^{(2)}_{max}=0.4526$; (c) $C(\rho^{(S)})$; (d)  $C(\rho^{(R)})$.}
}
  \label{Fig:c1c2nEq0} 
\end{figure*}

This case combines {the} properties of the two previous ones, {see Fig. \ref{Fig:c1c2nEq0}, but there are certain differences. Fig.\ref{Fig:c1c2nEq0}a and Fig.\ref{Fig:c1c2nEq0}c show that the concurrence $C(\rho^{(S)})$ for $c^{(1)}=0$ approximately repeats the shapes shown in Fig.\ref{Fig:c1Eq0}, while  $C(\rho^{(S)})$   for  $c^{(2)}=0$ is less than the entanglement in Fig.\ref{Fig:c2Eq0} and is zero for $N=42$}. There is a critical line $(c^{(1)}_{S;cr},c^{(2)}_{S;cr})$ on the plane $(c^{(1)},c^{(2)})$  which bounds the region with zero concurrence { (this  boundary is not depicted  explicitly in Fig.\ref{Fig:c1c2nEq0}a).}  The region of zero concurrence reduces with an increase in $N$. 

The concurrence $C(\rho^{(R)})$ behaves in a different way, Fig.\ref{Fig:c1c2nEq0}b,d. At $c^{(2)}=0$, it approximately repeats the shape shown in  
Fig.\ref{Fig:c2Eq0}, while it is constant { at $c^{(1)}=0$: $C(\rho^{(R)})|_{N=6}=0.1026$, $C(\rho^{(R)})|_{N=42}=0.0212$}.  
Thus $\rho^{(R)}$ is nonzero for all {positive} $c^{(1)}$ and $c^{(2)}$ from the ellipse-like domain with the semi-axes $c^{(1)}_{max}$ and $c^{(2)}_{max}$.
The Fig.\ref{Fig:c1c2nEq0}a and Fig.\ref{Fig:c1c2nEq0}c  show that the concurrence in the sender $C(\rho^{(S)})$ reaches larger values for a long chain than for a short one, unlike the concurrence in the receiver. {Comparing Fig.\ref{Fig:c1c2nEq0}a with Fig.\ref{Fig:c1c2nEq0}b (and Fig.\ref{Fig:c1c2nEq0}c with Fig.\ref{Fig:c1c2nEq0}d) we see that the maximum  value of $C(\rho^{(S)})$ 
 exceeds the  maximum  value of $C(\rho^{(R)})$ over  $(c^{(1)},c^{(2)})$. But all the sender's states with zero entanglement  are  mapped onto the receiver's states with nonzero entanglement.} 


\subsubsection{{ \bf Case IV:} $\lambda^{1}= \lambda^{2}$,
{$c^{(1)}\ge 0$, $c^{(2)} \ge 0$}.} 
\begin{figure*}
\subfloat[]{\includegraphics[scale=0.7,angle=0]{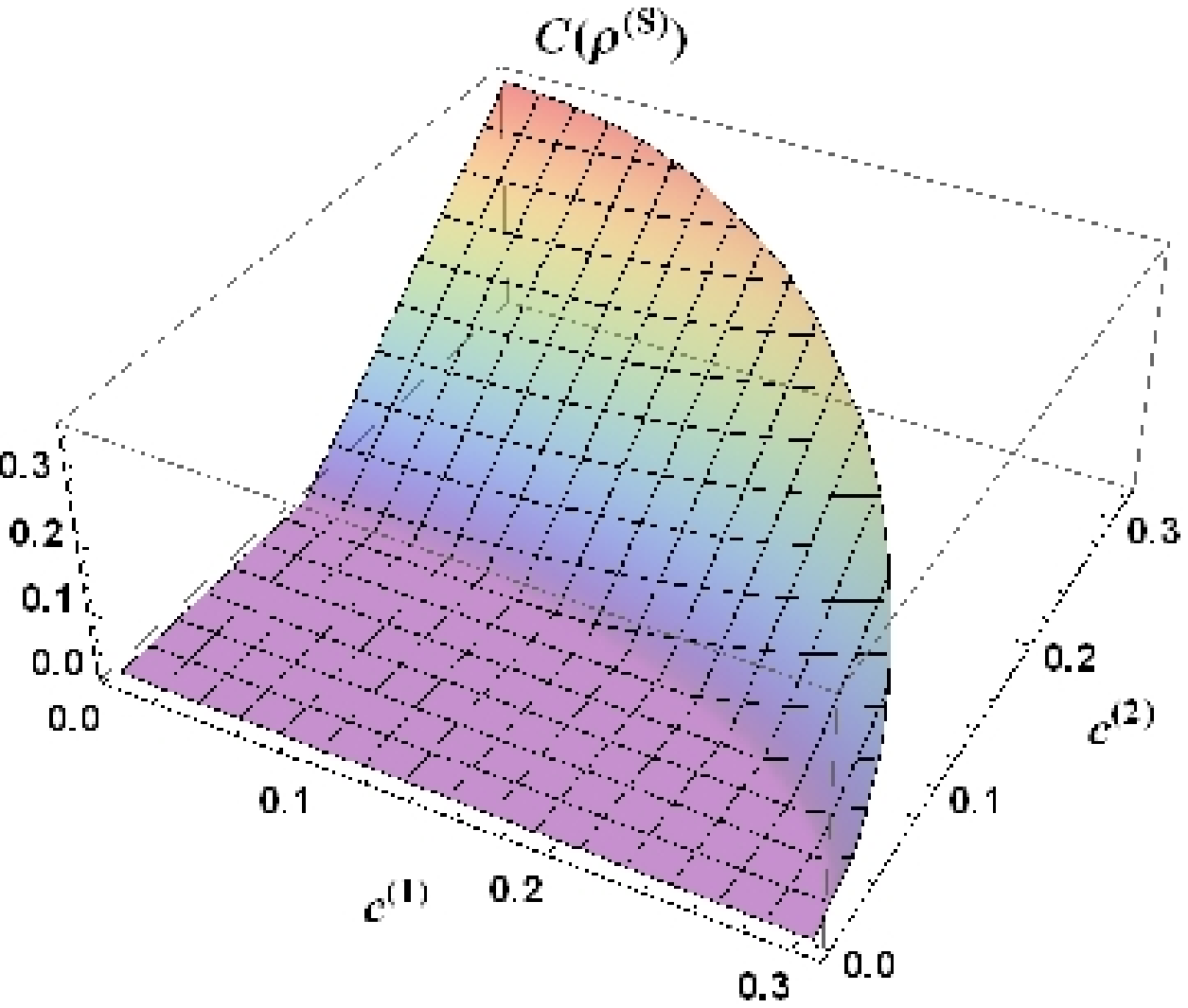
}}
\subfloat[]{\includegraphics[scale=0.7,angle=0]{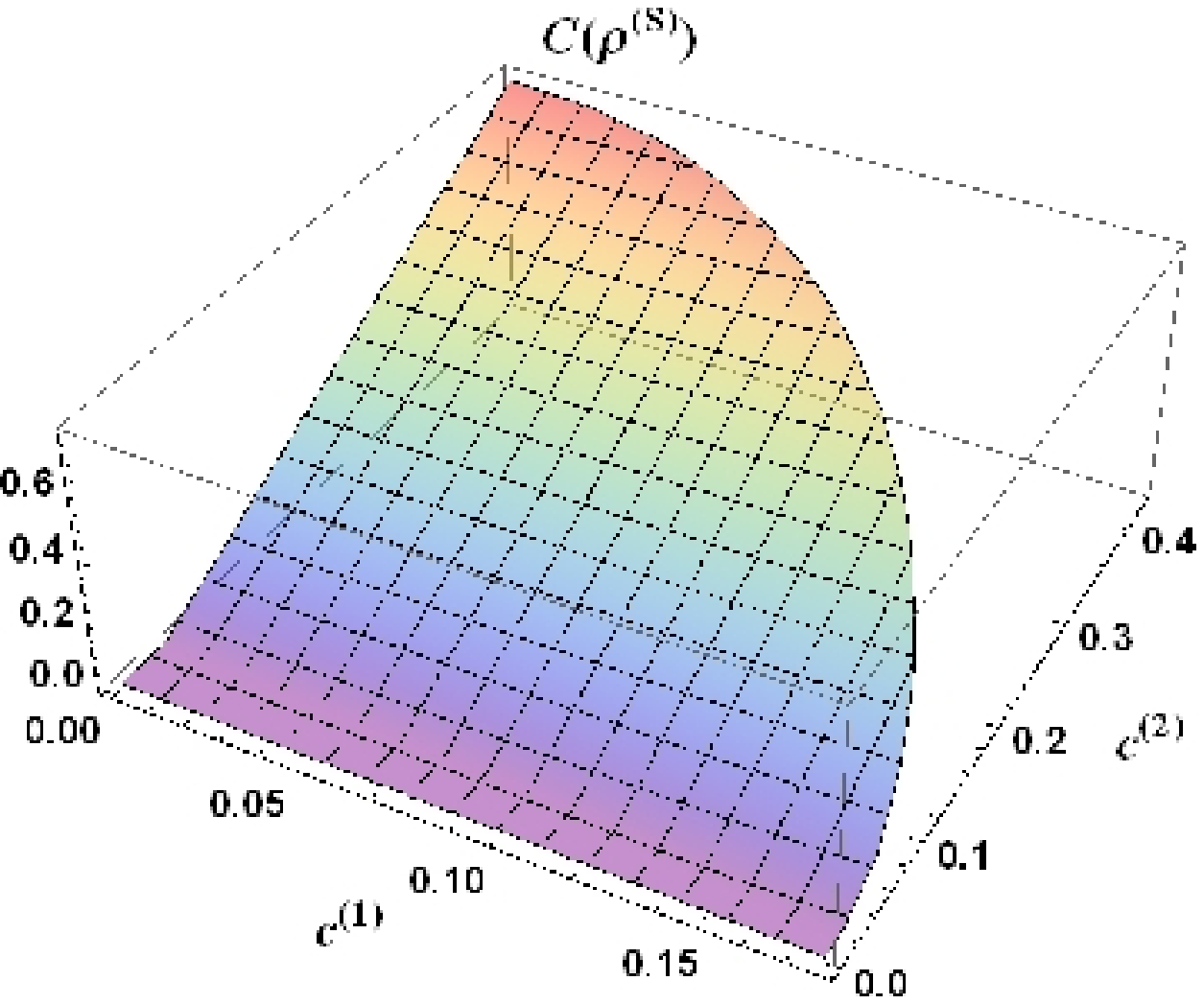
}}
\caption{{Case IV:} $\lambda^{(1)}= \lambda^{(2)}$, {$c^{(1)}\ge 0$, $c^{(2)}\ge 0$}.  Concurrence $C(\rho^{(S)})$ as a function of $c^{(1)}$
and $c^{(2)}$  over the quarter of the ellipse-like domain on the plane $(c^{(1)},c^{(2)})$. { (a) $N=6$, semi-axes $c^{(1)}_{max}=0.3058$ and $c^{(2)}_{max}=0.3200$;  (b) $N=42$, semi-axes
$c^{(1)}_{max}=0.1789$ and $c^{(2)}_{max}=0.4283$.}
}
  \label{Fig:lam1Eqlam2} 
\end{figure*}

Only the concurrence in $\rho^{(S)}$ is nonzero in this case. {The shapes of graphs} in  Fig.\ref{Fig:lam1Eqlam2}a,b  are similar to {that in} Fig.\ref{Fig:c1c2nEq0}a,c, but the values of the concurrences are  less in this case.


\section{Perturbations of block-scalable states and their influence on entanglement}

\label{Section:pert}

\subsection{Perturbations of block-scalable states}
\label{Section:pert1}

The two-qubit receiver's density matrix $\rho^{(R)}$  can be written in terms of the sender's density matrix
$\rho^{(S)}$ and the evolution operator $T$ as follows \cite{Z_JPA2012} 
\begin{eqnarray}\label{ev}
\rho^{(R)}_{nm}(t) = \sum_{i,j=1}^4 T_{nm,ij}(t)\rho^{(S)}_{ij}(0).
\end{eqnarray}
{Since, on the other hand,   $\rho^{(S)}$ and $\rho^{(R)}$  are representable  as, respectively,   sums (\ref{rhoSB}) and (\ref{rhoRB}), we conclude that} in the unperturbed case, the evolution operator $T$ reduces to multiplication of $\rho^{(S;i)}$ and $\tilde \rho^{(S;0)}$ by the scale factors. {In practice,} we can not construct 
matrices $\rho^{(X;i)}$ in expansions (\ref{rhoSB}) and (\ref{rhoRB}) exactly,
because of experimental and numerical errors. Therefore, the initial state   can deviate
from the required {block-scalable} state. { Consequently, the quantum correlations in such state can differ from those calculated in Sec. \ref{Section:conc}. This {prompts us to}  compare the quantum correlations in block-scalable (block-scaled) states with those in states from their neighborhood.} We introduce the perturbed MQ-coherence matrices $\rho^{(S;i)}(\varepsilon)$, $i=0,\pm 1, \pm2$:
\begin{eqnarray}
&&
\tilde \rho^{(S;0)}(\varepsilon) = \tilde \rho^{(X;0)} + \sigma^{(S;0)} \varepsilon,
\\\nonumber
&&
\rho^{(S;i)}(\varepsilon)+\rho^{(S;-i)}(\varepsilon) = (\rho^{(X;i)}+\rho^{(X;-i)} + (\sigma^{(S;i)}+\sigma^{(S;-i)}) \varepsilon) c^{(i)},\;\;i=1,2.
\end{eqnarray}
Here, $\varepsilon$ is the perturbation amplitude, $\sigma^{(S;i)}$ is a  perturbation of the $i$-order coherence matrix, $\sigma^{(S;0)}$ is a Hermitian matrix, ${\mbox{Tr}}\;\sigma^{(S;0)}=0$, and $(\sigma^{(S;i)})^+ =\sigma^{(S;-i)}$. In addition, $\sigma^{(S)}$ must be such that the perturbed matrix $\rho^{(S)}$ remains a density matrix 
({Hermitian non-negative} definite matrix with {the} unit trace) over the whole allowed domain in the plane $(c^{(1)},c^{(2)})$.
Thus, the perturbed initial state reads
\begin{eqnarray}
\rho^{(S)}(\varepsilon) = \rho^{(S)}(0) + \sigma^{(S)} \varepsilon,
\end{eqnarray}
where
\begin{eqnarray}
\sigma^{(S)}= \sigma^{(S;0)} + (\sigma^{(S;1)}+\sigma^{(S;-1)}) c^{(1)}+ (\sigma^{(S;2)}+\sigma^{(S;-2)}) c^{(2)}.
\end{eqnarray}
This results in the following perturbed
receiver's state:
\begin{eqnarray}
\rho^{(R)}_{nm}(\varepsilon) = \sum_{i,j=1}^4 T_{nm,ij}\rho^{(S)}_{ij}(0) + \sigma^{(R)} \varepsilon,
\end{eqnarray}
where
\begin{eqnarray}
\sigma^{(R)}_{nm}= \sum_{i,j=1}^4 T_{nm,ij}\sigma^{(S)}_{ij} \varepsilon.
\end{eqnarray}  
{We consider the random nondiagonal  elements} of $\sigma^{(S;k)}$, $k=0,1,2$, in the following form:
\begin{eqnarray}\label{pert:nond}
&&\sigma^{(S;k)}_{nm}=\sigma^{(k;1)}_{nm} e^{i \sigma^{(k;2)}_{nm}},\;\;k=0,1,2,\;\;n\neq m\\\nonumber
&&
0 \le \sigma^{(k;1)}_{nm}  \le 1,\;\;0 \le \sigma^{(k;2)}_{nm}  \le 2\pi.
\end{eqnarray}
To satisfy the normalization trace-condition for the diagonal elements we  perturb them as { follows}:
\begin{eqnarray}\label{pertd}
{
\sigma^{(S;0)}_{nn}=\tilde \sigma_{nn}-\frac{1}{4} \sum_{i=1}^4 \tilde \sigma_{ii},} \;\;
-1 \le \tilde \sigma_{ii}  \le 1.
\end{eqnarray}

\subsection{Effect of block-scalable state perturbation on entanglement}
\label{Section:num2}
Now we compare the concurrence of block-scalable and block-scaled states with the concurrence of the states from their  neighborhood. 
With this aim, we {consider} the effect of perturbation described in {Sec.\ref{Section:pert1}}  on the concurrence of both sender's and receiver's states {   averaging the calculated concurrence} over 5000 (Secs.\ref{Section:a}
and \ref{Section:b}) or 1000 (Secs.\ref{Section:c}
and \ref{Section:d})  realizations of an arbitrary matrix $\sigma^{(S)}$.
{Hereafter $C$ denotes this {mean} concurrence.} 
We consider { one of the following sets of six perturbation amplitudes in the communication lines of $N=6$ and $N=42$ nodes:}
\begin{eqnarray}
\label{vareps}
&&
\varepsilon =0.0125, \; 0.025, \;0.05, \;0.1, \;0.2,\;1,\;\;{\mbox{in Secs. \ref{Section:a} and \ref{Section:b} }},\\\nonumber
&&
\varepsilon =0.0125, \; 0.025, \;0.05, \;0.1, \;0.2,\;0.5,\;\;{\mbox{in Secs. \ref{Section:c} and \ref{Section:d} }}.
\end{eqnarray}
 {We include the large amplitude perturbations to explore the difference between the entanglements in the block-scalable (block-scaled) states and the entanglements in states which are far from them.} { Similar to Sec.\ref{Section:num}, we describe the entanglement in four cases indicated in the end of Sec.\ref{Section:block}.}

\subsubsection{{\bf Case I:} $\lambda^{1}\neq \lambda^{2}$, $c^{(1)}=0$, {$c^{(2)}\ge 0$}.}
\label{Section:a}

\begin{figure*}
\subfloat[]{\includegraphics[scale=0.6,angle=0]{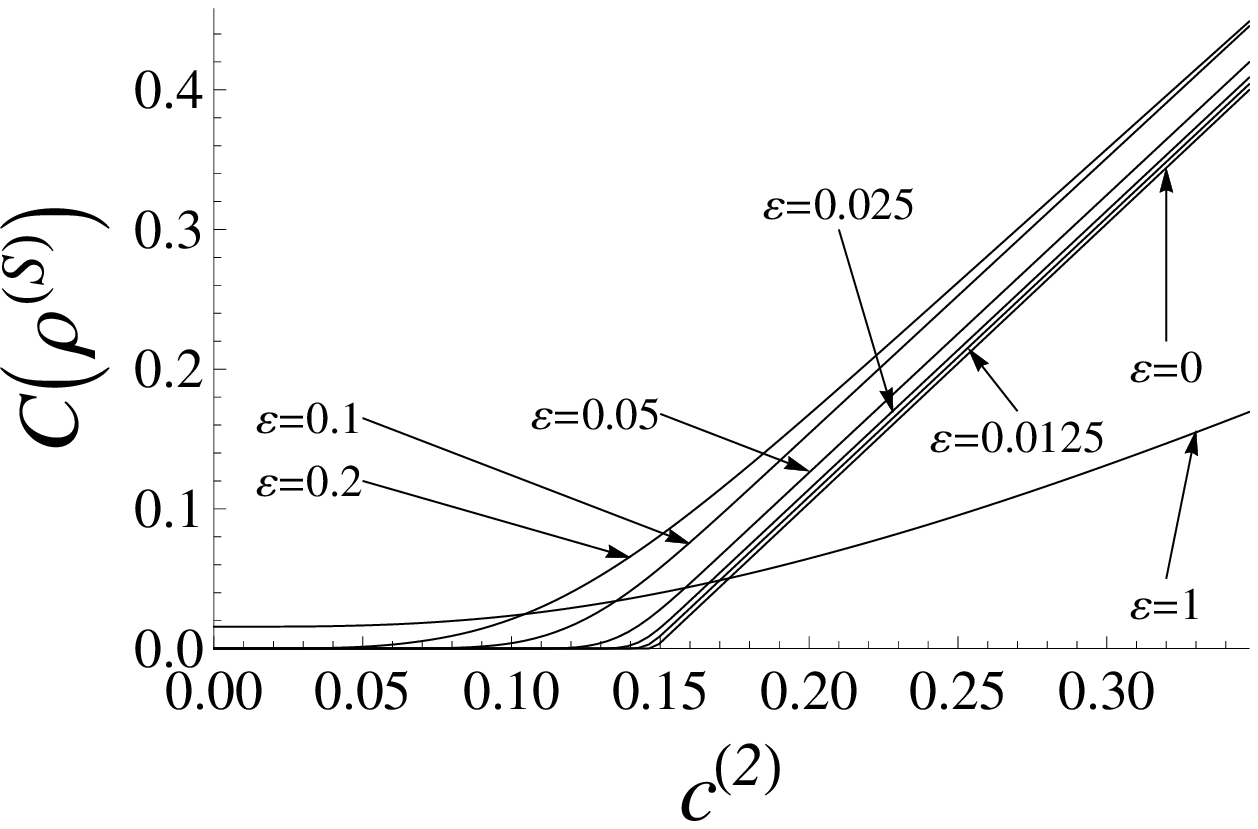
}}
\subfloat[]{\includegraphics[scale=0.6,angle=0]{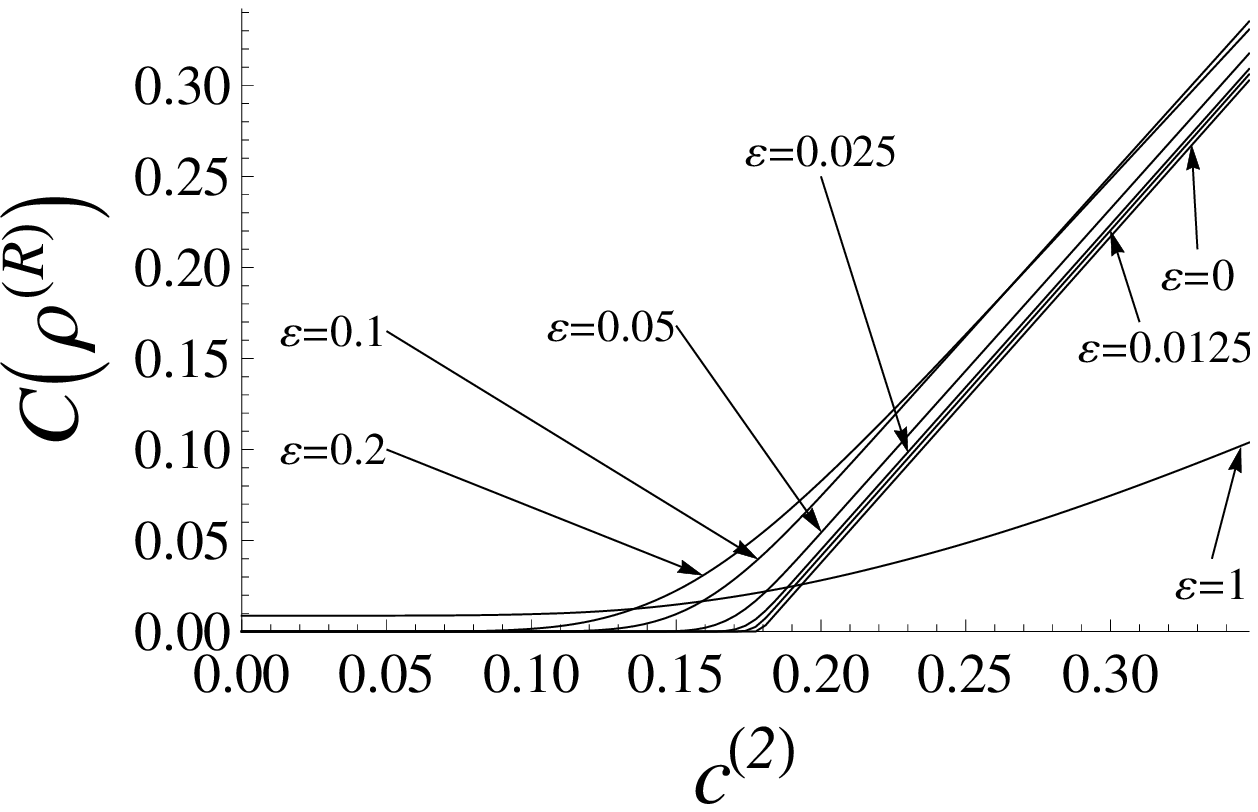
}}\\
\subfloat[]{\includegraphics[scale=0.6,angle=0]{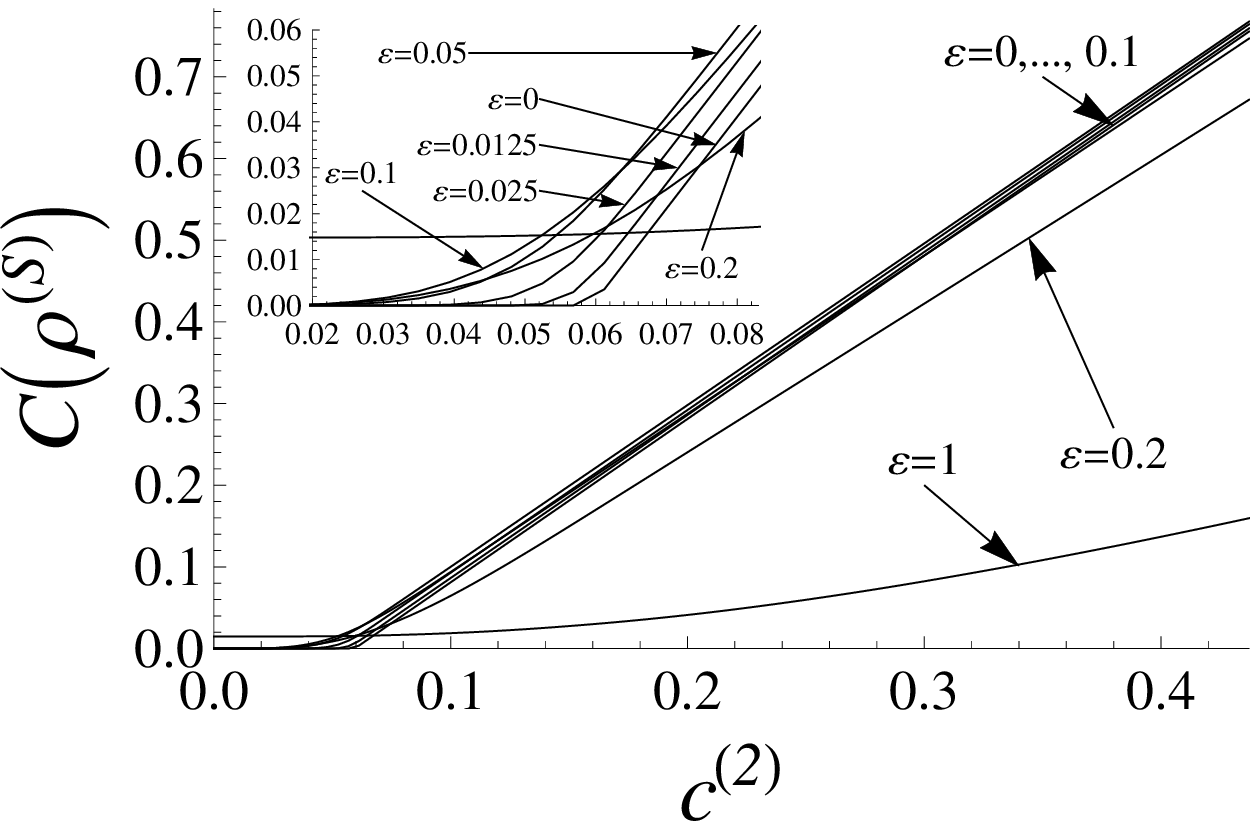
}}
\subfloat[]{\includegraphics[scale=0.6,angle=0]{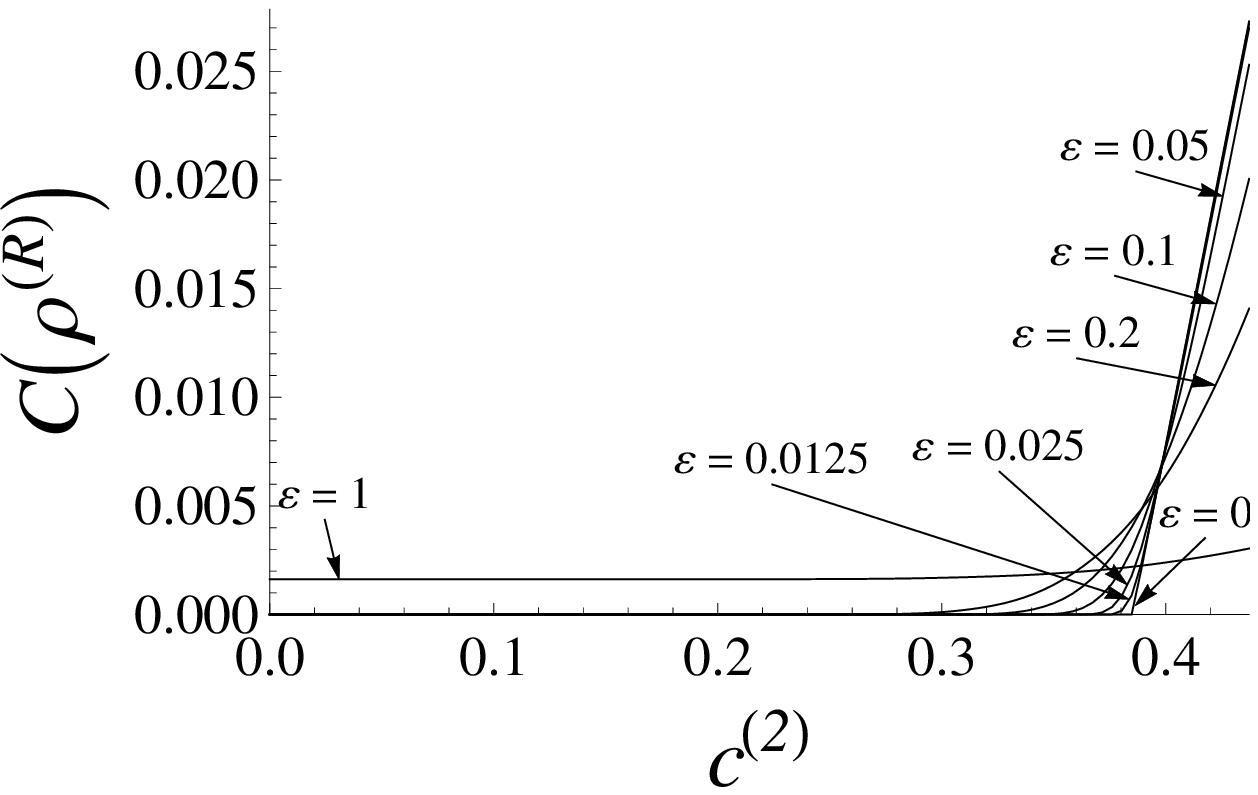
}}
\caption{\label{Fig:perta}{Case I:} $\lambda^{(1)}\neq\lambda^{(2)}$, $c^{(1)}= 0$, {$c^{(2)}\ge  0$}. Concurrences  $C(\rho^{(S)})$ and $C(\rho^{(R)})$ as functions of $c^{(2)}$ {for different values of the perturbation amplitude $\varepsilon$ (\ref{vareps}).  (a,b) $N=6$, $c^{(2)}_{max}=0.3479$; (a) $C(\rho^{(S)})$, the lines corresponding to $\varepsilon =0$, 0.0125, $\dots$, 0.2 form a bunch where $\varepsilon$ increases in an upward direction; (b) $C(\rho^{(R)})$, the lines corresponding to $\varepsilon = 0$, 0.0125, 0.025 form a bunch where $\varepsilon$ increases in  an upward direction.  (c,d) $N=42$, $c^{(2)}_{max}=0.4372$;
(c)  $C(\rho^{(S)})$, the  lines corresponding  to $\varepsilon =0$, 0.0125, 0.025, 0.05, 0.1 form a bunch; the line  $\varepsilon=0.1$ crosses all other lines in this bunch and becomes the lower one near $c^{(2)}_{max}$; other lines in this bunch do not cross each other, $\varepsilon$ of  these lines  increases in upward direction;  inset shows the behavior of the concurrences for small $c^{(2)}$; (d) $C(\rho^{(R)})$. 
}}
\end{figure*}

The perturbation of the sender's initial state  in this case reads
\begin{eqnarray}\label{ex1}
\sigma^{(S)} = \sigma^{(S;0)} + (\sigma^{(S;2)}+\sigma^{(S;-2)}) c^{(2)}.
\end{eqnarray}
The {mean} concurrences in $\rho^{(S)}(0)$ and $\rho^{(R)}$  {for}  set of $\varepsilon$ (\ref{vareps})    are shown in Fig.\ref{Fig:perta}.
This figure demonstrates that an increase in $\varepsilon$ leads to  a decrease in 
$c^{(2)}_{S;cr}$ and $c^{(2)}_{R;cr}$. But the maximal { value of
concurrence {$C_{max}(\rho^{(S)})$   (corresponding to $c^{(2)}_{max}$) { increases with an increase in  $\varepsilon$ till $\varepsilon \sim 0.2$ ($N=6$) or $\varepsilon \sim 0.05$ ($N=42$). But then  it decreases with the further  increase in $\varepsilon$}, see Fig.\ref{Fig:perta}a,c.}
Similarly,     $C_{max}(\rho^{(R)})$ in short chains increases with an increase in  $\varepsilon$ till $\varepsilon \sim 0.1$. Then, $C_{max}(\rho^{(R)})$ decreases with the further increase in $\varepsilon$, 
see Fig.\ref{Fig:perta}b. But in the long chain ($N=42$), $C_{max}(\rho^{(R)})$ decreases with an increase in $\varepsilon$ till $\varepsilon \sim 1$, see Fig.\ref{Fig:perta}d.}

\subsubsection{{\bf Case II:} $\lambda^{1}\neq \lambda^{2}$, {$c^{(1)}\ge 0$}, $c^{(2)} = 0$.}\label{Section:b}
\begin{figure*}
\subfloat[]{\includegraphics[scale=0.6,angle=0]{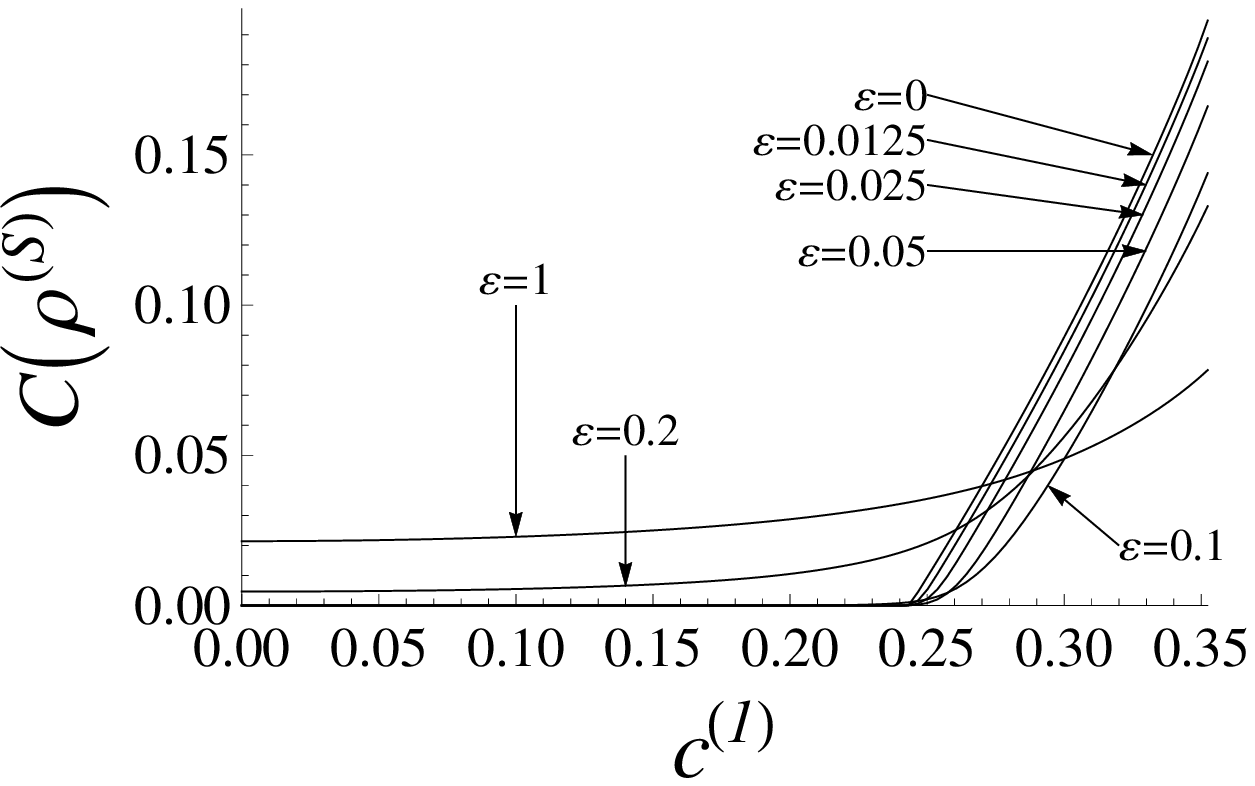
}}
\subfloat[]{\includegraphics[scale=0.6,angle=0]{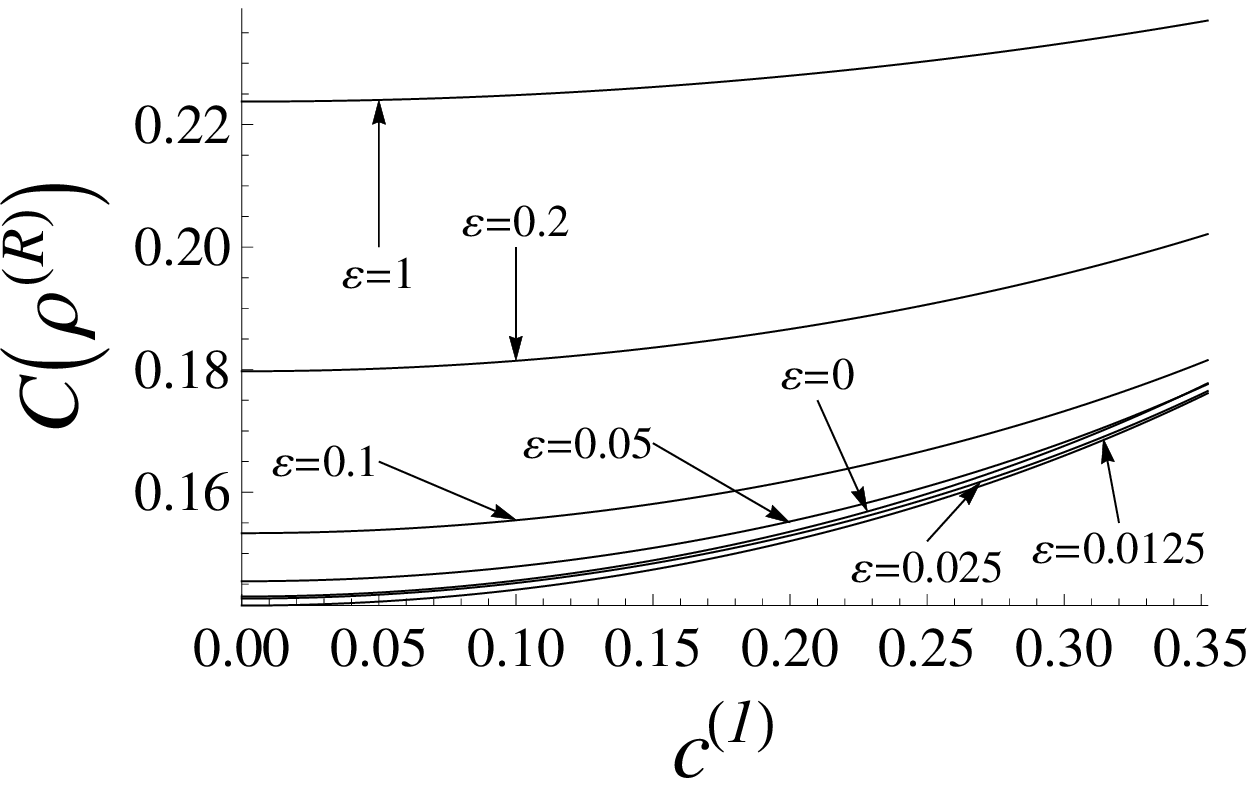
}}\\
\subfloat[]{\includegraphics[scale=0.6,angle=0]{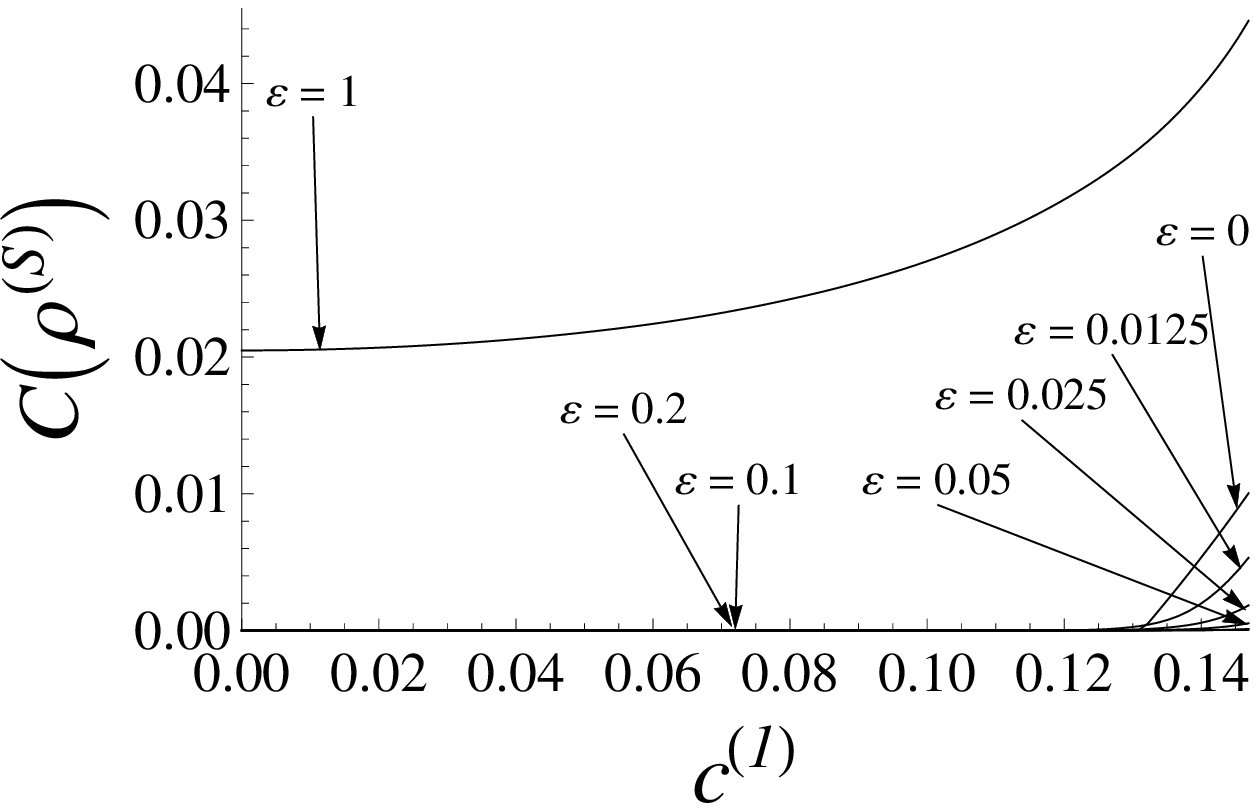
}}
\subfloat[]{\includegraphics[scale=0.6,angle=0]{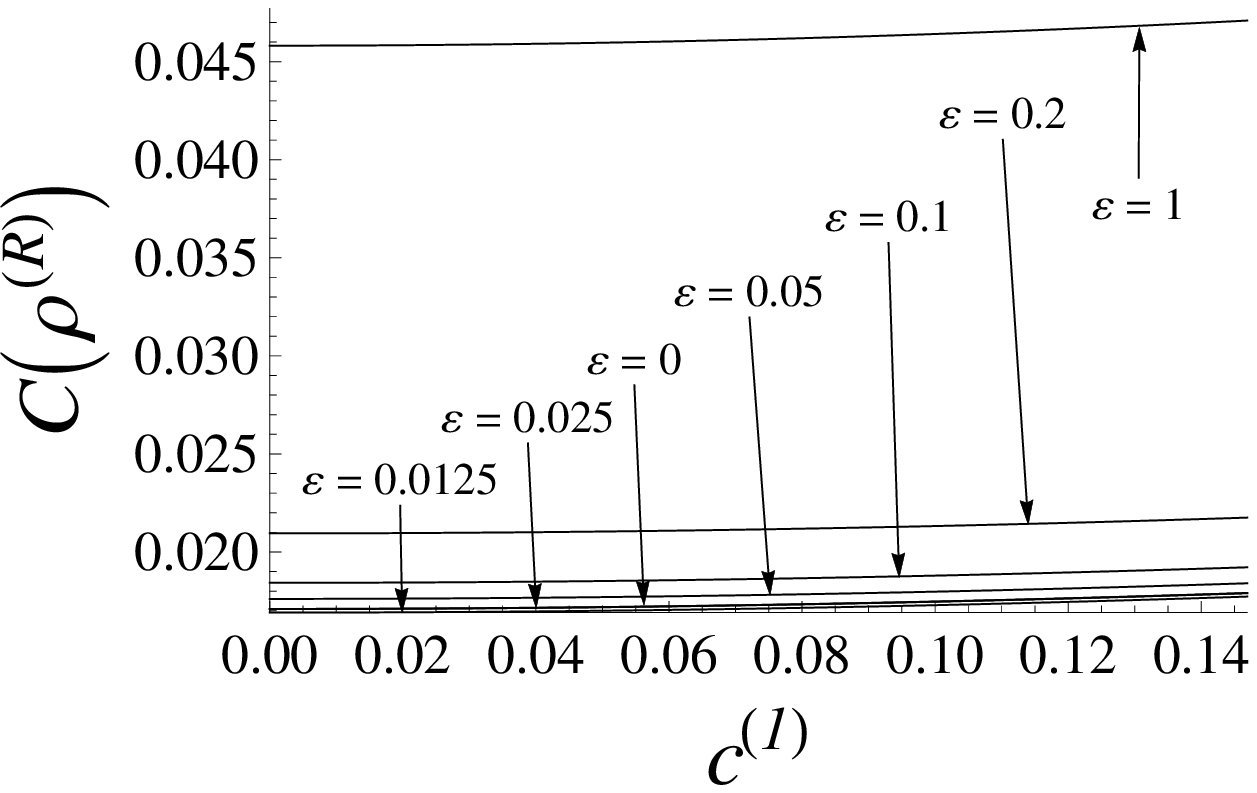
}}
\caption{ {Case II:} $\lambda^{(1)}\neq\lambda^{(2)}$, {$c^{(1)}\ge 0$}, $c^{(2)}= 0$. Concurrences $C(\rho^{(S)})$ and $C(\rho^{(R)})$ as functions of $c^{(1)}$ {for different values of the perturbation amplitude $\varepsilon$ (\ref{vareps}).  (a,b) $N=6$, $c^{(1)}_{max}=0.3524$; (a) $C(\rho^{(S)})$; (b) $C(\rho^{(R)})$, the lines $\varepsilon=0$ and $0.05$ cross each other.  (c,d) $N=42$, $c^{(1)}_{max}=0.1469$;
(c)  $C(\rho^{(S)})$, the lines corresponding to $\varepsilon=0.1$ and 0.2 are  indistinguishable from the axis of abscissas; at $c^{(2)}_{max}$, $C_{max}(\rho^{(S)})$ decreases with an increase in $\varepsilon$ till 
$\varepsilon\sim 0.2$ and increases after that; (d) $C(\rho^{(R)})$, the order of $\varepsilon$ is shown in figure. 
}}
  \label{Fig:pertb} 
\end{figure*}

The perturbation of the sender's initial state  in this case reads
\begin{eqnarray}\label{ex2}
\sigma^{(S)} = \sigma^{(S;0)} + (\sigma^{(S;1)}+\sigma^{(S;-1)}) c^{(1)} .
\end{eqnarray}
The mean {concurrences} for set of $\varepsilon$ (\ref{vareps}) are shown in Fig.\ref{Fig:pertb}. 
This figure {demonstrates} that an increase in $\varepsilon$ leads to  a decrease in $c^{(1)}_{S;cr}$. Also the maximal  
concurrence $C_{max}(\rho^{(S)})$ (corresponding to $c^{(1)}_{max}$) decreases with an increase in $\varepsilon$ in a short chain, see Fig. \ref{Fig:pertb}a. In the long chain ($N=42$), it decreases till $\varepsilon \sim 0.2$ and then it increases with an increase in $\varepsilon$, see Fig.\ref{Fig:pertb}c.
Behavior of $C(\rho^{(R)})$ is different, {see Fig.\ref{Fig:pertb}b and Fig.\ref{Fig:pertb}d. {First of all,} 
it is positive for all $c^{(1)}$. $C_{max}(\rho^{(R)})$  decreases with $\varepsilon$ till $\varepsilon \sim 0.0125$ in both short ($N=6$) and long ($N=42$) chains,  and then it increases with the further increase in $\varepsilon$.  For the long chain,  the
large deviation from the block-scalable state ($\varepsilon=1$) significantly increases the  concurrence in the sender and receiver for all $c^{(1)}$, {as shown in Fig.\ref{Fig:pertb}c,d. For the short chain, this holds only for $C(\rho^{(R)})$, Fig.\ref{Fig:pertb}b }}

\subsubsection{{\bf Case III:} $\lambda^{1}\neq \lambda^{2}$, {$c^{(1)}\ge 0$, $c^{(2)} \ge 0$}.}\label{Section:c}
\begin{figure*}
\subfloat[]{\includegraphics[scale=1.5,angle=0]{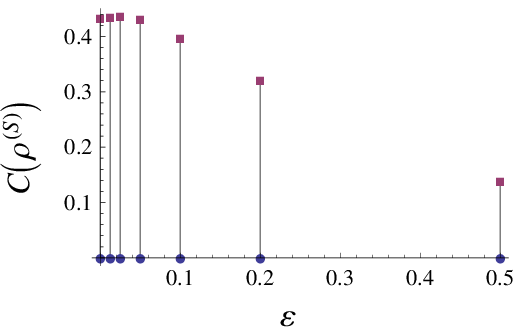
}}
\subfloat[]{\includegraphics[scale=1.5,angle=0]{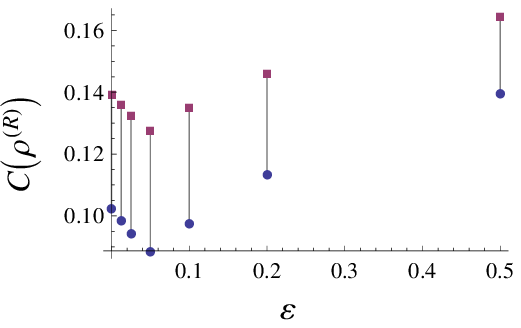
}}\\
\subfloat[]{\includegraphics[scale=1.5,angle=0]{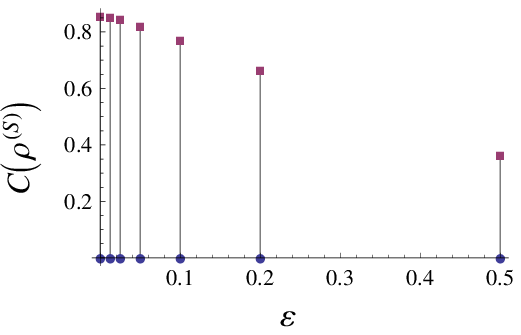
}}
\subfloat[]{\includegraphics[scale=1.5,angle=0]{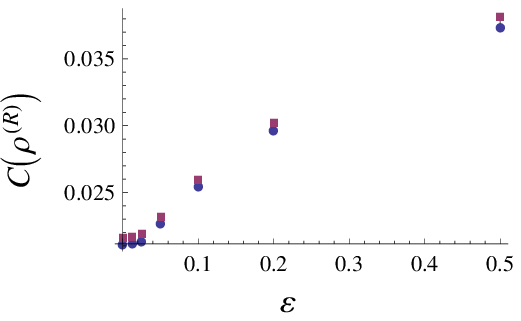
}}
\caption{{Case III:} $\lambda^{(1)}\neq\lambda^{(2)}$, {$c^{(1)}\ge  0$, $c^{(2)}\ge 0$}.  Maximal (squares) and minimal (circles) values of the concurrences $C(\rho^{(S)})$ and $C(\rho^{(R)})$  { (\ref{extr}) as functions of} $\varepsilon$.   (a) $N=6$, $C(\rho^{(S)})$; (b) $N=6$, $C(\rho^{(R)})$; (c) $N=42$, $C(\rho^{(S)})$; (d) $N=42$, $C(\rho^{(R)})$.
}
  \label{Fig:c1c2} 
\end{figure*}

The perturbation of the sender's initial state  in this case reads
\begin{eqnarray}\label{ex3}
\sigma^{(S)} = \sigma^{(S;0)} + (\sigma^{(S;1)}+\sigma^{(S;-1)}) c^{(1)}+ (\sigma^{(S;2)}+\sigma^{(S;-2)}) c^{(2)}.
\end{eqnarray}
{We characterize the effect of perturbations by} the extrema { $C_{max} =\displaystyle \min_{c^{(1)},c^{(2)}} C$ and $C_{min} =\displaystyle \max_{c^{(1)},c^{(2)}} C$} as functions of $\varepsilon$ in Fig.\ref{Fig:c1c2}.
{ The numerical study shows that the functions $C(\rho^{(S)})$ and $C(\rho^{(R)})$ take their maximal and minimal values at the same
points on the plane $(c^{(1)},c^{(2)})$ for all perturbations:}
\begin{eqnarray}\label{extr}
&&
C_{min}(\rho^{(S)}(0))=\min_{c^{(1)},c^{(2)}} C(\rho^{(S)}(0)) =C(\rho^{(S)}(0))|_{{c^{(1)}=0}\atop{c^{(2)}=0}},\\\nonumber
&&
C_{max}(\rho^{(S)}(0))=\max_{c^{(1)},c^{(2)}} C(\rho^{(S)}(0)) =C(\rho^{(S)}(0))|_{{c^{(1)}=0}\atop{c^{(2)}=c^{(2)}_{max}}},\\\nonumber 
&&
C_{min}(\rho^{(R)})=\min_{c^{(1)},c^{(2)}} C(\rho^{(R)}) =C(\rho^{(R)})|_{{c^{(1)}=0}\atop{c^{(2)}=0}},\\\nonumber
&&
C_{max}(\rho^{(R)})=\max_{c^{(1)},c^{(2)}} C(\rho^{(R)}) =C(\rho^{(R)})|_{{c^{(1)}=c^{(1)}_{max}}\atop{c^{(2)}=0}}. 
\end{eqnarray}
Figs.\ref{Fig:c1c2}a and \ref{Fig:c1c2}c {demonstrate} that the minimal value of $C(\rho^{(S)})$ equals zero for all $\varepsilon$, while {$C_{max}(\rho^{(S)})$ in the short chain ($N=6$) has a weakly formed maximum at $\varepsilon\sim 0.025$, and then it} decreases with an increase in $\varepsilon$.  The behavior of  $C(\rho^{(R)})$ is different, Fig.\ref{Fig:c1c2}b,d. Its minimal value is never zero and both minimal and maximal values generally increase with an increase in $\varepsilon$ for large $\varepsilon$. {In the short chain of $N=6$ nodes, both $C_{min}(\rho^{(R)})$ and $C_{max}(\rho^{(R)})$ have a minimum at  $\varepsilon \sim 0.05$}. We also notice   
that the minimal and maximal values of $C(\rho^{(R)})$ approach each other in long chains, as shown in  Fig.\ref{Fig:c1c2}d. In other words, the concurrence tends to the constant value in this case.

\subsubsection{{\bf Case IV:} $\lambda^{1}= \lambda^{2}$, {$c^{(1)}\ge 0$, $c^{(2)} \ge 0$}.}
\label{Section:d}
\begin{figure*}
\subfloat[]{\includegraphics[scale=1.5,angle=0]{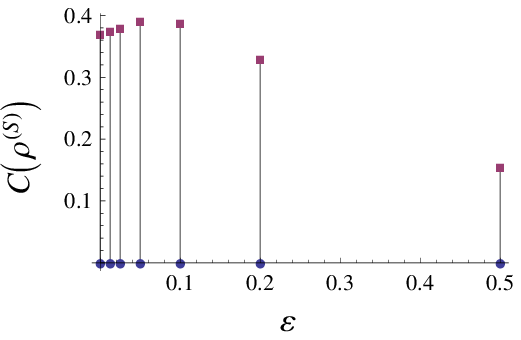
}}
\subfloat[]{\includegraphics[scale=1.5,angle=0]{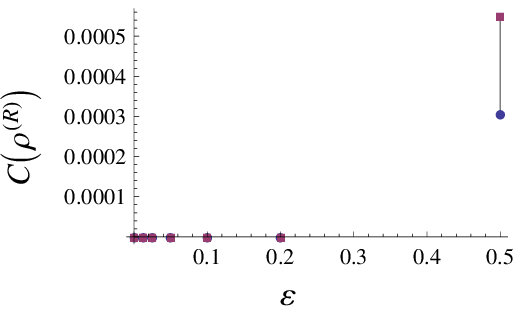
}}\\
\subfloat[]{\includegraphics[scale=1.5,angle=0]{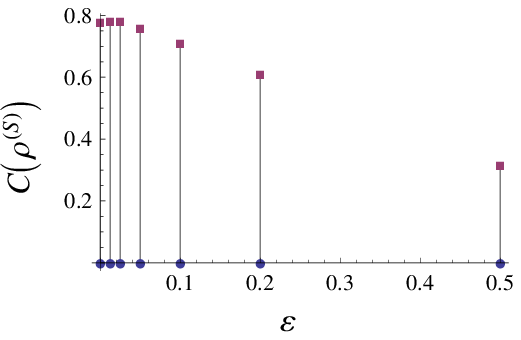
}}
\subfloat[]{\includegraphics[scale=1.5,angle=0]{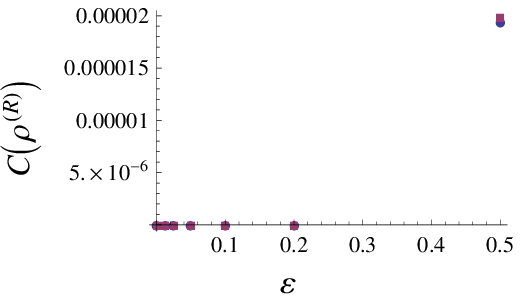
}}
\caption{{Case IV:} $\lambda^{(1)}=\lambda^{(2)}$, {$c^{(1)}\ge  0$, $c^{(2)}\ge 0$}.  Maximal (squares) and minimal (circles) values of the concurrences $C(\rho^{(S)})$ and $C(\rho^{(R)})$  { as functions of} $\varepsilon$.   (a) $N=6$, $C(\rho^{(S)})$; (b) $N=6$, $C(\rho^{(R)})$; (c) $N=42$, $C(\rho^{(S)})$; (d) $N=42$, $C(\rho^{(R)})$.
}
  \label{Fig:c1c2l1l2} 
\end{figure*}

The perturbation of the sender's initial state  in this case is as in (\ref{ex3}).
The extrema  $\displaystyle  C_{min}(\rho^{(S)})$ and $\displaystyle C_{max}(\rho^{(S)})$ (\ref{extr}) as functions of $\varepsilon$ are shown in  Fig.\ref{Fig:c1c2l1l2}. We see that the behavior of the maximal and minimal values of 
$C(\rho^{(S)})$ {in Fig.\ref{Fig:c1c2l1l2}a,c is very similar  to that in  Case III, Sec.\ref{Section:c}}.
The only difference {is the presence of a {well formed} maximum at $\varepsilon\sim 0.05$ in the short chain of $N=6$ spins and a weakly formed maximum at $\varepsilon \sim 0.0125$ in the long chain.} On the contrary,  $C(\rho^{(R)})$ is nonzero only for large $\varepsilon$
{($\varepsilon \sim 0.5$)}. In this case, again {the}
minimal and maximal values of $C(\rho^{(R)})$ approach each other {in the long chain with $N=42$, i.e, the  concurrence  is nearly independent on  $c^{(1)}$ and $c^{(2)}$  for all perturbation  amplitudes $\varepsilon$, see 
Fig.\ref{Fig:c1c2l1l2}d.}

From the analysis of Cases I -- IV in Secs.\ref{Section:a}-\ref{Section:d} we  conclude that in the {framework} of our consideration, the {maximal value of} concurrence in the block-scalable { sender's initial states, $\displaystyle C_{max}(\rho^{(S)}(0))$,}
reaches larger values than that in the  states from the remote neighborhood {($\varepsilon = 1$ in Cases I, II and $\varepsilon = 0.5$ in Cases III, IV ), except in the case $c^{(2)}=0$, $N=42$ (long chain), 
Fig.\ref{Fig:pertb}c}. However, the corresponding concurrence  {$\displaystyle C_{max}(\rho^{(R)})$} in the receiver's block-scaled states  is smaller 
than that in the states from the remote neighbourhood   (except the Case I, Fig.\ref{Fig:perta}b,d).
 {This is the price to pay for the simple and well-described deformation of the transferred quantum state.} 
Regarding the uniform scaling in {Case IV, Sec.\ref{Section:d} ($\lambda^{(1)}=\lambda^{(2)}$)}, the concurrence in the block-scaled states does not appear at all in the settings of our numerical simulations except the case of large-amplitude perturbations {($\varepsilon=0.5$), Fig.\ref{Fig:c1c2l1l2}b,d.}

 \section{Conclusions}
 \label{Section:conclusions}
The block-scaled state transfer is a method {for} remote state creation such that the state created at the receiver is similar to the initial sender's state  with a minimal, well-described deformation. This deformation appears {as}  the  scale factors {in front of}  certain blocks (MQ-coherence matrices) of the initial sender's state  except the only diagonal element which must satisfy the trace  condition. {In this case the initial sender's state is referred to as the block-scalable state.}

{Since the map (block-scalable state) $\to$ (block-scaled  state) can be simply described, it is interesting to find out features of quantum correlations in such states.  These correlations are studied in our paper in terms of entanglement (concurrence) { in spin chains with 2-qubit sender and receiver}.
We find out the dependence  of the concurrence  on the  parameters $c^{(1)}$ and $c^{(2)}$ transferred by the $\pm 1$- and $\pm 2$-order coherence matrices}. There are certain differences in concurrences for the four cases outlined  in the end of Sec.\ref{Section:block}. { Thus, in Case I, the concurrence appears only for large $c^{(2)}$ in both $\rho^{(S)}$ and $\rho^{(R)}$.  
In  Case II, the concurrence in  $\rho^{(S)}$ appears  for large $c^{(1)}$, while the concurrence in $\rho^{(R)}$ appears for all $c^{(1)}$. Case III combines properties of Cases I and II, { but $C(\rho^{(R)})$ is nonzero for all values of $c^{(1)}$ and $c^{(2)}$ from 
their  domain.}  Case IV differs from others by having zero concurrence in the receiver's state. Generally, { the concurrence in the block-scalable states reaches  significantly larger values  than the concurrence in the block scaled states.}

{Next, we estimate the difference  between the entanglement in the block-scalable (block-scaled) states and in the states from their close and remote neighborhoods. To this end, }
we study the effect of perturbations of the initial block-scalable states  on the entanglement and show that this effect  depends on which of the higher order coherence matrices are involved in the process. 
Thus, if only the zeroth- and $\pm2$-order coherences are transferred {(Case I)}, then,  
for the small perturbations, entanglement  appears only for large enough $c^{(2)}$ in both $\rho^{(S)}$ and $\rho^{(R)}$. But eventually for $\varepsilon \sim 1$, entanglement appears for all $c^{(2)}$ and its maximal value is significantly smaller then that for the unperturbed case. If only zeroth- and $\pm1$-order coherences are used {(Case II)}, then the entanglement {$C(\rho^{(S)})$} appears for large $c^{(1)}$ if $\varepsilon\ll 1$. But, for $\varepsilon \sim 1$, this entanglement appears for all $c^{(1)}$ and, in long chains, it becomes significantly larger than the entanglement in the unperturbed case.  The entanglement of the receiver state  { $C(\rho^{(R)})$} exists  for all $c^{(1)}$ and it generally increases (except the very small perturbations amplitudes $\varepsilon$) with an increase in $\varepsilon$. {In both these  cases,} the critical values $c^{(1)}_{S;cr}$,  $c^{(1)}_{R;cr}$ and $c^{(2)}_{S;cr}$  {decrease } with an increase in $\varepsilon$.   If all three coherence matrices are involved {(Case III)}, then the maximal value of entanglement as a function of $c^{(1)}$ and $c^{(2)}$ generally decreases with an increase in $\varepsilon$ in the sender, while it generally increases in the receiver.
 If $\lambda^{(1)}=\lambda^{(2)}$ {(Case IV)}, then the entanglement in the sender behaves similar to the previous case, and the entanglement in the receiver $C(\rho^{(R)})$ becomes non-zero only for large-amplitude perturbations. { In  Cases II-IV applied to a long chain ($N=42$),
the entanglement in the receiver $C(\rho^{(R)})$  only slightly depends on the parameters $c^{(1)}$ and $c^{(2)}$.}

{Thus,  the block-scaled state transfer is a variant of a state evolution with a simple and well-described state deformation. Usually, the entanglement of the receiver's block-scaled state  is less than the entanglement in the corresponding sender's block-scalable state. The entanglement in the block-scaled state can even vanish.}
{Large perturbation of a block-scalable state leads to a decrease in concurrence  $C(\rho^{(S)})$ and {to} an increase in concurrence $C(\rho^{(R)})$}.
{In addition,  the evolution generally  reduces the dependence of concurrence on the parameters $c^{(1)}$ and $c^{(2)}$ {in long chains}, which is referred to both unperturbed and perturbed cases considered in this paper. }

This work was performed in accordance with the state task, state 
 registration No. 0089-2019-0002, and was  partially supported by the program of the Presidium of RAS No. 5 ''Photonic technologies in probing inhomogeneous media and biological objects''.
 
\section{Appendix: Explicit form of block-scalable ($\rho^{(S)}(0)$) and block-scaled ($\rho^{(S)}(t)$) density matrices}
\label{Section:AppendixA}
We present the density matrices used in the numerical calculations of Sec.\ref{Section:num}. More details can be found in 
Ref.\cite{BFZ_Arch2018} where these matrices are constructed  as a result of the optimization procedure. 
In formulas (\ref{case11})-(\ref{case42}) below we use the following notation: 
\begin{eqnarray}
\Lambda^{(0)}=\left\{
\begin{array}{cc}
1,& \rho^{(B)}=\rho^{(S)}\cr
\lambda^{(0)},& \rho^{(B)}=\rho^{(R)}
\end{array}
\right.,\;\;
{\Lambda^{(n)}=\left\{
\begin{array}{cc}
c^{(n)},& \rho^{(B)}=\rho^{(S)}\cr
c^{(n)}\lambda^{(n)},& \rho^{(B)}=\rho^{(R)}
\end{array}
\right.,\;\;n=1,2.}
\end{eqnarray}
\begin{enumerate}
\item
$\lambda^{1}\neq \lambda^{2}$,
$c^{(1)}=0$, {$c^{(2)}\ge 0$}, $\lambda^{(0)}(6)=1.08371$, $\lambda^{(2)}(6)=0.89602$, $t(6)=8.51533$,  $b(6)=b(42)=10$,
$\lambda^{(0)}(42)=1.69754$, $\lambda^{(2)}(42) =0.26204$, $t(42)=47.97194$,
%
\begin{eqnarray}\label{case11}
\rho^{(B)}(6)=\left(
\begin{array}{cccc}
0.40596 \Lambda^{(0)}& 
    0& 0 & \Lambda^{(2)}\cr
0&     0.15131 \Lambda^{(0)}& 
     0.00010 i \Lambda^{(0)}& 0\cr 
     0& - 0.00010 i \Lambda^{(0)}& 
    0.14467 \Lambda^{(0)}&0 \cr 
    \Lambda^{(2)}& 0 &0 & 
   1-0.70194\Lambda^{(0)}
\end{array}\right).
\end{eqnarray}

\begin{eqnarray}\label{case12}
\rho^{(B)}(42)=\left(
\begin{array}{cccc}
0.44635\Lambda^{(0)}&0&0&\Lambda^{(2)}\cr
0&0.08266 \Lambda^{(0)}&0.01664 i \Lambda^{(0)}&0\cr
0&-0.01664 i \Lambda^{(0)}&0.04275 \Lambda^{(0)}&0\cr
\Lambda^{(2)}&0&0&1-0.57176 \Lambda^{(0)}
\end{array}\right).
\end{eqnarray}

\item
$\lambda^{1}\neq \lambda^{2}$,
{$c^{(1)}\ge 0$}, $c^{(2)} = 0$, $\lambda^{(0)}(6)=1.22015$, $\lambda^{(1)}(6)=0.81452$,
$t(6)=5.03255$, $b(6)=10$,  $\lambda^{(0)}(42)=1.36938$, $\lambda^{(1)}(42) =0.31866$,  $t(42)=41.32805$, $b(42)=7.02476$,

\begin{eqnarray}\label{case21}
\rho^{(B)}(6)=\left(
\begin{array}{cccc}
0.59440 \Lambda^{(0)}&
    0.77790 \Lambda^{(1)}& - 
     0.62839 i \Lambda^{(1)}& 0\cr
     0.77790 \Lambda^{(1)}&
    0.12890 \Lambda^{(0)}& 
     - 0.10707 i \Lambda^{(0)}&  - 
     0.00003 i \Lambda^{(1)}\cr
     0.62839 i \Lambda^{(1)}& 0.10707 i \Lambda^{(0)}& 
    0.09232 \Lambda^{(0)}& -0.00004 \Lambda^{(1)}\cr
    0& 
     0.00003 i \Lambda^{(1)}& -0.00004 \Lambda^{(1)} & 1-0.81562 \Lambda^{(0)}
\end{array}\right).
\end{eqnarray}

\begin{eqnarray}\label{case22}
\rho^{(B)}(42)=\left(
\begin{array}{cccc}
0.68654 \Lambda^{(0)} &0.92494 \Lambda^{(1)}&-0.38010 i \Lambda^{(1)}&0\cr
0.92494 \Lambda^{(1)}&0.03320\Lambda^{(0)}&-0.01710 i \Lambda^{(0)}&-0.00034 i \Lambda^{(1)}\cr
0.38010 i \Lambda^{(1)}&0.01710 i \Lambda^{(0)}&0.01035 \Lambda^{(0)}&-0.00082 \Lambda^{(1)}\cr
0& 0.00034 i \Lambda^{(1)}&-0.00082 \Lambda^{(1)}&1-0.73008 \Lambda^{(0)}
\end{array}\right).
\end{eqnarray}

\item
$\lambda^{1}\neq \lambda^{2}$,
{$c^{(1)}\ge 0$, $c^{(2)} \ge 0$}, $\lambda^{(0)}(6)=1.26340 $, $\lambda^{(1)}(6)=0.76126$, $\lambda^{(2)}(6)=0.22885$, $t(6)=5.37677$, $b(6)=5.37902$,
$\lambda^{(0)}(42)=1.53227$, $\lambda^{(1)}(42) =0.29521$, $\lambda^{(2)}(42) =0.03935$, $t(42)=41.94097$,  $b(42)=6.87959$,

\begin{eqnarray}\label{case31}
\rho^{(B)}(6)=\left(
\begin{array}{cccc}
0.51945\Lambda^{(0)}&
    0.88361 \Lambda^{(1)}& - 
     0.46820 i \Lambda^{(1)}& \Lambda^{(2)}\cr
     0.88361 \Lambda^{(1)}& 
    0.18237\Lambda^{(0)}&  - 0.11342 i\Lambda^{(0)}& - 
     0.00216 i \Lambda^{(1)}\cr
     0.46820 i \Lambda^{(1)}&  0.11342 i\Lambda^{(0)}& 
    0.07949\Lambda^{(0)}& -0.00408 \Lambda^{(1)}\cr
    \Lambda^{(2)}& 
     0.00216 i \Lambda^{(1)}& -0.00408 \Lambda^{(1)} &1- 0.78130\Lambda^{(0)}
\end{array}\right).
\end{eqnarray}

\begin{eqnarray}\label{case32}
\rho^{(B)}(42)=\left(
\begin{array}{cccc}
0.58895 \Lambda^{(0)} &0.96932 \Lambda^{(1)}&-0.24580 i \Lambda^{(1)}&\Lambda^{(2)}\cr
0.96932 \Lambda^{(1)}&0.05117 \Lambda^{(0)}&-0.02255 i \Lambda^{(0)}&-0.00025 i \Lambda^{(1)}\cr
0.24580i \Lambda^{(1)}&0.02255 i \Lambda^{(0)}&0.01209 \Lambda^{(0)}&-0.00010 \Lambda^{(1)}\cr
\Lambda^{(2)}&0.00025 i \Lambda^{(1)}&-0.00010 \Lambda^{(1)}&1-0.65221 \Lambda^{(0)}
\end{array}\right).
\end{eqnarray}
\item
$\lambda^{1} = \lambda^{2}$, {$c^{(1)}\ge 0$, $c^{(2)} \ge 0$}, $\lambda^{(0)}(6)=1.2022$, $\lambda^{(1)}(6)=\lambda^{(2)}(6)=0.2956$, $t(6)=5.6651$, $b(6)=2.3462$,
$\lambda^{(0)}(42)=1.42581$, $\lambda^{(1)}(42) =\lambda^{(2)}(42) =0.04943$, 
 $t(42)=42.3077$,  $b(42)=3.8922$,
\begin{eqnarray}\label{case41}
\rho^{(B)}(6)=\left(
\begin{array}{cccc}
0.49962 \Lambda^{(0)} & 
  0.98333 \Lambda^{(1)}&  - 0.15484 i \Lambda^{(1)}& 
  \Lambda^{(2)} \cr
  0.98333 \Lambda^{(1)}& 
  0.20645 \Lambda^{(0)}  &  - 
   0.07298 i \Lambda^{(0)} &- 
   0.01482 i \Lambda^{(1)}\cr
   0.15484 i \Lambda^{(1)}& 
   0.07298 i \Lambda^{(0)} & 
  0.08884  \Lambda^{(0)} & -0.09414 \Lambda^{(1)} \cr
  \Lambda^{(2)}& 
   0.01482 i \Lambda^{(1)}& -0.09414 \Lambda^{(1)}& 
  1-0.79491 \Lambda^{(0)} 
\end{array}\right).
\end{eqnarray}
\begin{eqnarray}\label{case42}
\rho^{(B)}(42)=\left(
\begin{array}{cccc}
0.61003 \Lambda^{(0)}&0.99758 \Lambda^{(1)}&-0.06640 i \Lambda^{(1)}&\Lambda^{(2)}\cr
0.99758 \Lambda^{(1)}&0.06660 \Lambda^{(0)}&-0.02143 i \Lambda^{(0)}&-0.00135 i \Lambda^{(1)}\cr
0.06640 i \Lambda^{(1)}&0.02143 i \Lambda^{(0)}&0.02265 \Lambda^{(0)}&-0.02035 \Lambda^{(1)}\cr
\Lambda^{(2)}&0.00135 i \Lambda^{(1)}&-0.02035 \Lambda^{(1)}&1-0.69928 \Lambda^{(0)}
\end{array}\right).
\end{eqnarray}
\end{enumerate}

\end{document}